\documentclass[twocolumn,showpacs,preprintnumbers,amsmath,amssymb]{revtex4}
\usepackage{dcolumn}
\usepackage{bm}
\usepackage{graphicx}
\usepackage{amsmath}
\begin{document}

\title{Second order tunneling of two interacting bosons in a driven triple well}
\author{Zheng Zhou$^{1,2}$, \ Wenhua Hai$^{1}$\footnote{Corresponding author. Email address:
whhai2005@yahoo.com.cn}, \ Qiongtao Xie$^{1}$, \ Jintao Tan$^{1}$}
\affiliation{$^{1}$Department of Physics and Key Laboratory of
Low-dimensional Quantum Structures and\\ Quantum Control of Ministry
of Education, Hunan Normal University, Changsha 410081, China\\
$^{2}$Department of Physics and Mathematics, Hunan Institute of Technology, Hengyang 421002, China}

\begin{abstract}

We investigate quantum tunneling of two repulsive bosons in a triple-well potential
subject to a high-frequency driving field. By means of the multiple-time-scale asymptotic analysis, we evidence a far-resonant strongly-interacting
regime in which the selected coherent destruction of tunneling can occur between the paired states and unpaired states, and the
dominant tunneling of the paired states is a second order process. Two Floquet
quasienergy bands of the both kinds of states are given analytically, where a fine structure up to the second order corrections is displayed. The analytical results are
confirmed numerically based on the exact model, and may be particularly relevant to controlling correlated tunneling in experiments.

\end{abstract}


\maketitle

\section{Introduction}

Advances in laser technology have enabled studies of quantum tunneling and its coherent control for a single particle
in light-induced quantum wells without dissipation \cite{Oliver}. Research attempting to
manipulate quantum states has been underway for a long time
\cite{Grifoni, Kral}.
The time-periodic driving field is a powerful tool to control the tunneling dynamics and
can lead to important phenomena, such as dynamic localization (DL) \cite{Dunlap, Dignam}, coherent
destruction of tunneling (CDT) \cite{Grossmann,Kierig}, and photon-assisted tunneling \cite{Eckardt200401, Esmann,Xie2010}.
In recent years, the effects of interparticle interaction have attracted much attention. It was shown
that adjusting the interaction can give rise to richer behavior, including many-body selective CDT \cite{Gong, Longhi2012many}
and the second order tunneling of two interacting bosons \cite{Folling}. The two-body interaction model is the simplest model for studying the interacting effects, and has received much attention \cite{Liang, Wang, Kuo, Sascha},
since the seminal experimental result was reported \cite{Winkler}. The tunneling dynamics is related to the interplay between the
interparticle interaction and external field, and the former can be tuned by the
Feshbach resonance technique \cite{Zwierlein}.

In the presence of interaction and periodic external field, the quantum well system may be nonintegrable that necessitates the perturbation method for an analytical investigation. The multiple-scale technique
is a very useful perturbation method and has been extensively employed for different physical
systems \cite{Watanabe,Yan1996, Huang2008, Xie2009,Longhi2008}. It was demonstrated that with the multiple-scale perturbation method, the usual high-frequency approximation corresponds to the first-order perturbation correction \cite{Longhi2008}. In the far-resonant strongly-interacting regime with a stronger reduced interaction \cite{Creffield}, the high-frequency approximation is no longer valid. In this case, the dominant tunneling of paired states is a second-order process of long time scale and it can be described by the second-order perturbation correction. The correlated tunneling of two strongly-interacting atoms corresponding to time-resolved second
order tunneling has been observed directly in an undriven double well system \cite{Folling}. Very recently,
Longhi et al studied the second order effect of two far-resonant strongly-interacting bosons
in a periodically driven optical lattice by using a multiple-time-scale asymptotic analysis \cite{Longhi2012two}.

As above-mentioned, a lot of works on tunneling dynamics of two interacting atoms focus on the systems with double-well
or optical lattice potentials.
The triple-well system is a bridge between the double well and the optical lattice
systems, and is very important for us to fully understand coherent control of particle
tunneling in the quantum wells \cite{Bergmann,Lahaye, Gengbiao, Bin, Lushuai, Bradly}. Besides, the triple-well system itself owns some novel phenomena, e.g., the stimulated Raman
adiabatic passage \cite{Bergmann}, which is a scheme that adiabatically transport a quantum particle from
the left well to the right well with negligible middle well occupation at all times. The
tunneling dynamics of two-particle in a triple-well system have also attracted extensive attention \cite{Kirsten, Schneider,LuG}, however, research on the second order effect of the system has not been reported yet.

In this paper, we investigate the coherent control of the second order tunneling for two triple-well confined bosons driven by a high-frequency
laser field. By means of the multiple-time-scale asymptotic analysis, we characterize quantum dynamics of the two bosons with the continuous increase of interaction intensity, and demonstrate a far-resonant strongly-interacting regime in which two bosons initially occupying the same well would form a stable bound pair, because of the selected CDT between the paired states and the unpaired states. Taking into account the second order tunneling effect, the prediction on the CDT
is confirmed by the Floquet quasienergy analysis, where the Floquet quasienergy band of the three
unpaired states exhibits the avoided level-crossings (or new level-crossings) at (or near) the collapse points, and the fine structure of quasienergy band of the three paired states shows the different level-crossings beyond the former collapse points. Good agreements between the analytical and numerical results are shown, which could be verified further under
the current accessible experimental setups \cite{Winkler, Folling,Schlagheck}.

\section{The model and high-frequency approximation}

We consider two interacting bosons confined in a triple-well potential and driven
by an ac field. The Hamiltonian of the system in the tight-binding approximation
is described by the three-site Bose-Hubbard model \cite{Kirsten, Schneider,LuG}
\begin{eqnarray}\label{eq1}
\hat{H}(t)=&-&J(\hat{a}_1^\dag\hat{a}_{2}+\hat{a}_{2}^\dag\hat{a}_1+\hat{a}_2^\dag\hat{a}_{3}+\hat{a}_{3}^\dag\hat{a}_2)
\nonumber\\&+&\frac{U_0}{2}\sum\limits_{l=1}^3\hat{a}_l^\dag \hat{a}_l^\dag \hat{a}_l \hat{a}_l
+\varepsilon(t)(\hat{a}_3^\dag\hat{a}_3-\hat{a}_1^\dag\hat{a}_1),
\end{eqnarray}
where the operator $\hat{a}_l^{(\dag)}$ annihilates (creates) a boson in well $l$;  $J$ denotes
the nearest-neighbor hopping matrix element, $U_0$ is the on-site interaction energy,
and $\varepsilon(t)=\varepsilon\cos(\omega t)$ is the ac driving of amplitude $\varepsilon$ and frequency $\omega$.
For simplicity, we adopt $\hbar=1$ throughout this paper. The reference frequency $\omega_0=100$Hz is used to
normalize the energy and the parameters $J$, $U_0$, $\varepsilon$ and $\omega$, and
time $t$ is normalized in units of $\omega_0^{-1}$ such that all the quantities become dimensionless \cite{Holthaus,Hai}. Here we have assumed that the
three wells are deep enough such that the Wannier functions of the two interacting bosons belonging
to different wells have very small overlap. A Fock basis $|N_L, N_M, N_R\rangle$ is useful to describe
the two interacting bosons in the triple-well system, where $N_L$, $N_M$ and $N_R$ are the number
of bosons localized in the left, middle and right wells, respectively, with $N_L+N_M+N_R=2$. The
quantum state $|\psi(t)\rangle$ of the system is expanded as the linear superposition
of the Fock states,
\begin{eqnarray}\label{eq2}
|\psi(t)\rangle &=c_1(t)|2,0,0\rangle+c_2(t)|0,2,0\rangle+c_3(t)|0,0,2\rangle+\nonumber\\&c_4(t)|1,1,0\rangle
+c_5(t)|1,0,1\rangle+c_6(t)|0,1,1\rangle,
\end{eqnarray}
where $c_j(t)$ ($j=1,2,...,6$) denote the time-dependent probability amplitudes of finding the two bosons in
the six different Fock states and they obey the normalization condition $\sum_{j=1}^6 |c_j(t)|^2=1$.
Inserting Eqs. (1) and (2) into Schr\"{o}dinger equation $i\partial_t|\psi(t)\rangle=\hat{H}(t)
|\psi(t)\rangle$, one obtains the coupled equations for the amplitudes $c_j(t)$
\begin{eqnarray}\label{eq3}
&&i\dot{c}_1=[U_0-2\varepsilon(t)]c_1-\sqrt{2}Jc_4,\nonumber\\
&&i\dot{c}_2=U_0c_2-\sqrt{2}J(c_4+c_6),\nonumber\\
&&i\dot{c}_3=[U_0+2\varepsilon(t)]c_3-\sqrt{2}Jc_6,\nonumber\\
&&i\dot{c}_4=-\varepsilon(t)c_4-J(\sqrt{2}c_1+\sqrt{2}c_2+c_5),\nonumber\\
&&i\dot{c}_5=-J(c_4+c_6),\nonumber\\
&&i\dot{c}_6=\varepsilon(t)c_6-J(\sqrt{2}c_2+\sqrt{2}c_3+c_5).
\end{eqnarray}
Although it is difficult to obtain exact analytical solutions
of Eq. (3), we can approximately study some interesting phenomena in the
high-frequency regime with $\omega\gg J$. To do so, we rewrite the interaction strength as $U_0=m\omega+u$ for $|u|\le \omega/2,\ m=0,1,2,...$ with $u$ being the reduced interaction strength \cite{Creffield}, and make the function transformations $c_1(t)=a_1(t)\exp[-iU_0t+2i\varphi(t)]$,
$c_2(t)=a_2(t)\exp(-iU_0t)$, $c_3(t)=a_3(t)\exp[-iU_0t-2i\varphi(t)]$,
$c_4(t)=a_4(t)\exp[i\varphi(t)]$, $c_5(t)=a_5(t)$, and $c_6(t)=a_6(t)
\exp[-i\varphi(t)]$, with $a_j(t)$ being the slowly-varying functions and $\varphi(t)=\int_0^t \varepsilon
\cos(\omega\tau)d\tau=\frac{\varepsilon}{\omega}\sin(\omega t)$. Then, Eq.
(3) is transformed into the coupled equations in terms of $a_j(t)$. Under the
high-frequency approximation, the rapidly oscillating functions included in
the equations can be replaced by their time average such that the equations of $a_j(t)$
become \cite{LuG}
\begin{eqnarray}\label{eq4}
i\dot{a}_1=&-&\sqrt{2}J\mathcal{J}_{m}(\frac{\varepsilon}{\omega})a_4e^{iut},\nonumber\\
i\dot{a}_2=&-&\sqrt{2}J\Big[(-1)^m\mathcal{J}_{m}(\frac{\varepsilon}{\omega})a_4+\mathcal{J}_{m}(\frac{\varepsilon}{\omega})a_6\Big]e^{iut},\nonumber\\
i\dot{a}_3=&-&\sqrt{2}J(-1)^m\mathcal{J}_{m}(\frac{\varepsilon}{\omega})a_6e^{iut},\nonumber\\
i\dot{a}_4=&-&J\Big[\sqrt{2}\mathcal{J}_{m}(\frac{\varepsilon}{\omega})a_1+(-1)^m\sqrt{2}\mathcal{J}_{m}(\frac{\varepsilon}{\omega})a_2\Big]e^{-iut}\nonumber\\&-&J\mathcal{J}_{0}(\frac{\varepsilon}{\omega})a_5,\nonumber\\
i\dot{a}_5=&-&J\mathcal{J}_{0}(\frac{\varepsilon}{\omega})(a_4+a_6),\nonumber\\
i\dot{a}_6=&-&J\Big[\sqrt{2}\mathcal{J}_{m}(\frac{\varepsilon}{\omega})a_2+(-1)^m\sqrt{2}\mathcal{J}_{m}(\frac{\varepsilon}{\omega})a_3\Big]e^{-iut}\nonumber\\&-&J\mathcal{J}_{0}(\frac{\varepsilon}{\omega})a_5,
\end{eqnarray}
where $\mathcal{J}_{m}$ is the $m$th-order Bessel function of the first kind, and $e^{\pm iut}$ are the slowly varying functions for a small $u$ value. 

In Ref. \cite{Longhi2008}, Longhi proposed that the well-known high-frequency approximation
commonly used to study CDT corresponds to the first-order perturbation
approximation of the multiple-time-scale asymptotic analysis. If the first-order
correction term vanishes in the perturbation treatment, the
high-order corrected terms become important. Noticing that for a set of fixed external field parameters
dynamical behavior of the system (\ref{eq4}) is related to the self-interaction intensity.
In this work, we do not concern about the very strong interaction
(e.g., $U_0\geq 6\omega$), since for such a interaction we need to consider not only the usual on-site atom-interaction
strength, but also the interactions between atoms on neighboring lattice sites \cite{liang033617}, which is beyond the considered case.

When the condition
$\mathcal{J}_{0}=0$ is satisfied, Eq. (\ref{eq4}) shows that CDT occurs for the weakly-interacting case ($m=0,\ |u|\ll \omega$) that leads all the first derivatives of the probability amplitudes to zero. This can be further confirmed by
calculation of the Floquet quasienergies of the system in Sec. \uppercase\expandafter{\romannumeral 4}.
Besides, for the resonant strongly-interacting case ($u=0, m=1,2,...$), CDT for paired states
is observed when the condition $\mathcal{J}_{m}=0$ is satisfied that leads the first derivatives $\dot a_j(t)\ (j=1,2,3)$ of paired-state amplitudes to zero. This is consistent
with that of two interacting electrons in quantum dot arrays by numerical
computation of the Floquet quasienergies \cite{Creffield2004}. As an example, we show time evolutions of the
probabilities $P_j(t)=|c_j|^2=|a_j|^2 (j=1,2,...,6)$ for the resonant case with $J=1$,
$U_0=\omega=80$, $\varepsilon/\omega=2.405$ and $P_2(0)=1,\ P_{j\ne 2}(0)=0$, as in Fig. 1, where the first order result (the circular points) from Eq. (4) is confirmed by the direct numerical simulation (the curves) of
Eq. (3). From Fig. 1, we can see that transitions between the paired states with probabilities $P_j,\ j=1,2,3$ in Fig. 1(a) and the unpaired states with probabilities  $P_j,\ j=4,6$ in Fig. 1(b) happen periodically for the resonant case. We will come back to this property for comparison with the difference from the far-resonant case in next section.
\begin{figure}[htp] \center
\includegraphics[width=1.6in]{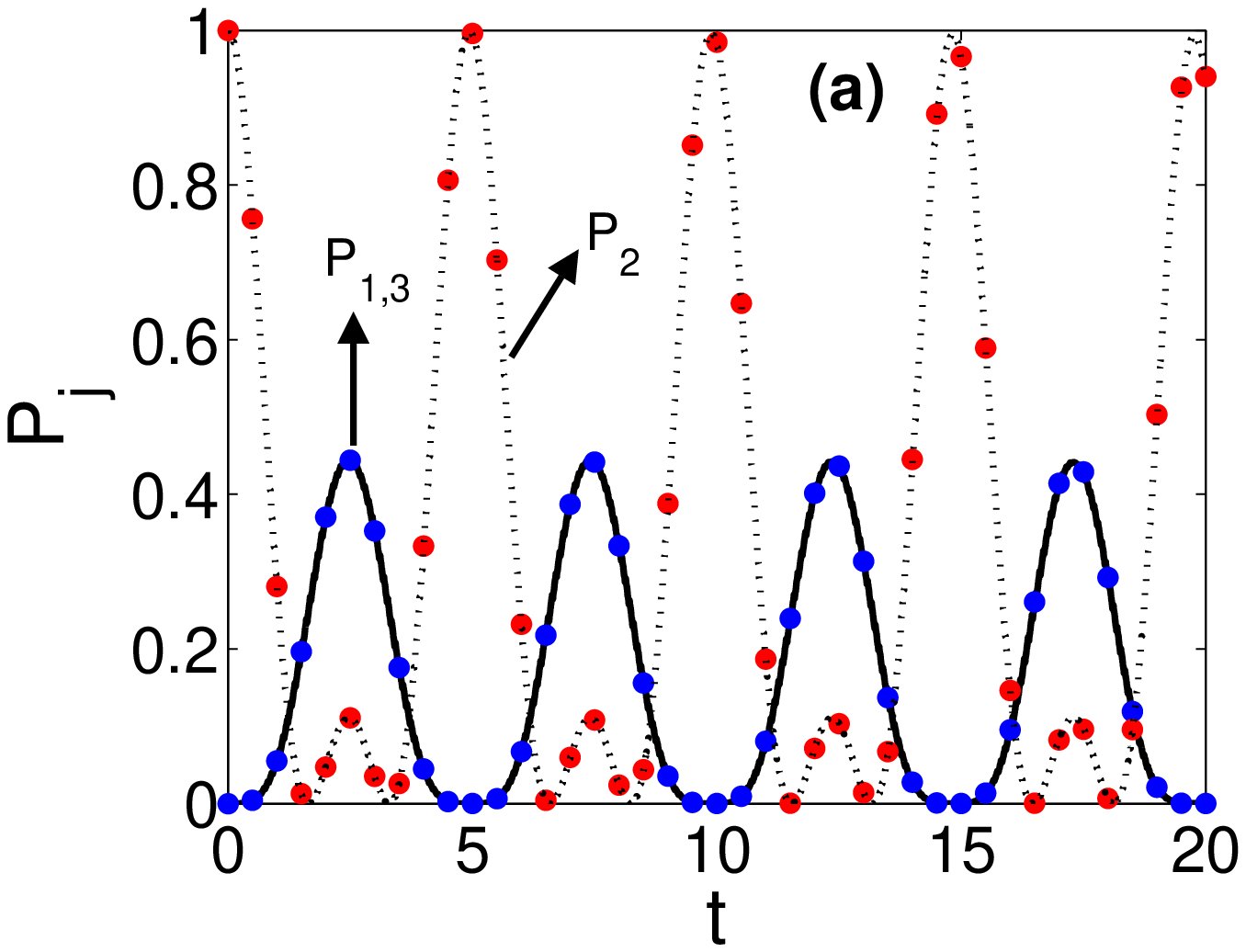}
\includegraphics[width=1.6in]{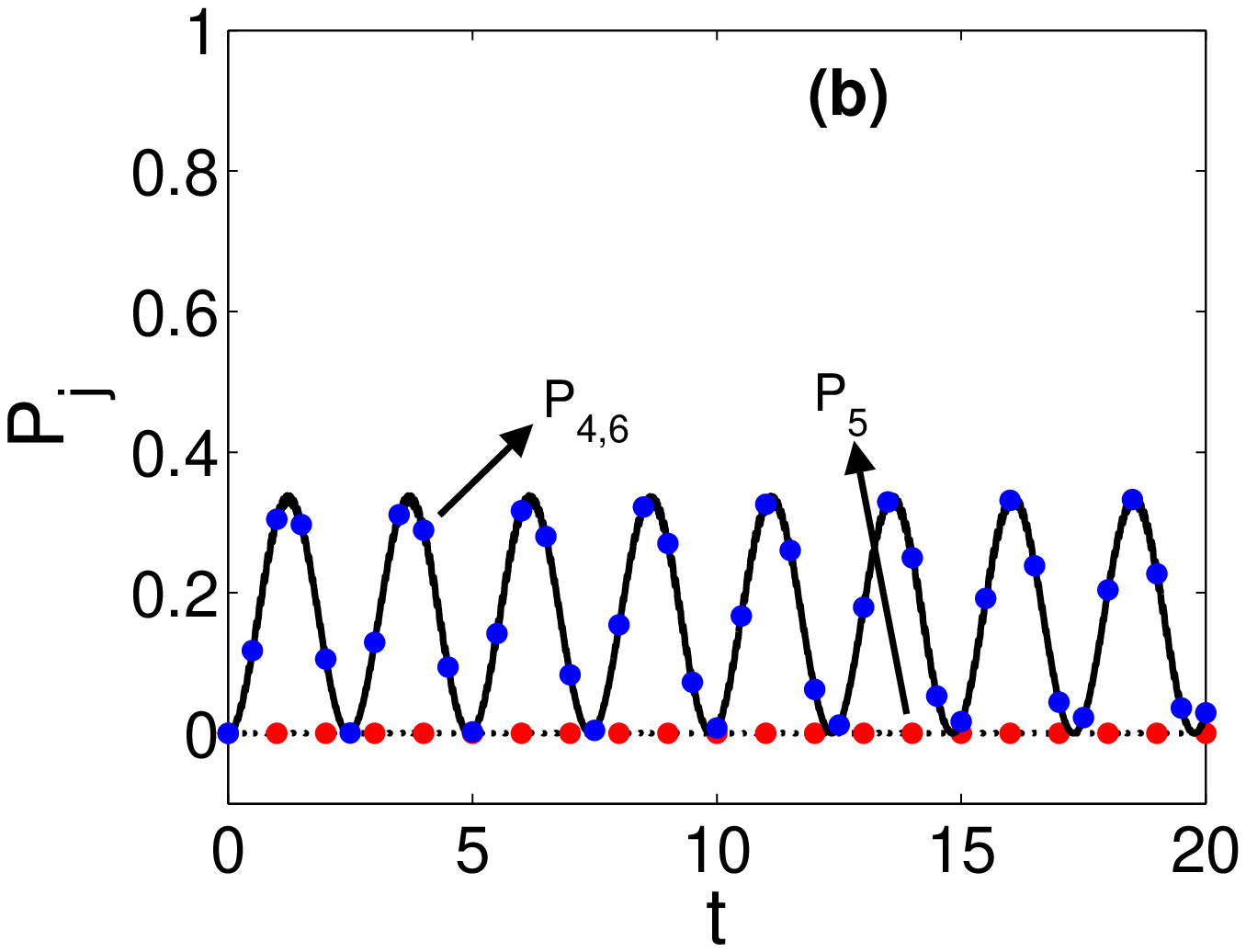}
\caption{\scriptsize{(Color online) Time evolutions of the probabilities $P_j=|a_j|^2
(j=1,2,...,6)$ in six different states for two bosons initially occupying
the middle well. The parameters are set as $J=1$, $U_0=\omega=80$, $\varepsilon/\omega
=2.405$. (a) The probabilities of the three paired states, where the dashed line corresponds to $P_2$, and the solid line to $P_{1,3}$.
(b) The probabilities of the three unpaired states, where the dashed line associates with $P_5$, and the solid line with $P_{4,6}$.
The circular points indicate the numerical results from the first approximate Eq. (4) and the curves describe the numerical solutions of the original Eq. (3).
Hereafter, all variables and parameters are dimensionless.}}
\end{figure}

We have known that Eq. (4) is a good approximation of Eq. (3) only for small values of the reduced interaction strength, $|u|\ll \omega$. When the $|u|$ values tend to their maximum $|u|= \omega/2$, the functions $e^{\pm iut}$ vary middlingly fast compared to the rapidly oscillating driving field. Consequently, in Eq. (4), although $e^{\pm iut}$ may be replaced by their average value of zero \cite{LuG} such that probability amplitudes $a_1(t), a_2(t)$ and $a_3(t)$ of the paired states are frozen approximately, effectiveness of the high-frequency approximation is lost partly. Particularly, in the case of moderate $|u|$ values, namely the values are neither very small nor too large, Eq. (4) is no longer a good approximation. Anyhow, for a stronger reduced interaction with a larger $|u|$ value, we require to employ other approximation methods and to explore the second order tunneling effects.

\section{second-order tunneling in the far-resonant strongly-interacting regime}

Now we consider the far-resonant case with a stronger reduced interaction to investigate tunneling dynamics of the system,
by means of multiple-time-scale asymptotic analysis.
In the high-frequency regime, we set $\epsilon=J/\omega$ as a small positive
parameter and $t'=\omega t$ is the rescaling time. The probability amplitudes $a_j(t')$ ($j=1,2,...,6$) are expanded
as a power series of $\epsilon$
\begin{equation}\label{eq5}
a_j(t')=a_j^{(0)}(t')+\epsilon a_j^{(1)}(t')+\epsilon^2 a_j^{(2)}(t')+\cdot\cdot\cdot.
\end{equation}
Owing to the high order infinitesimal can be neglected in the high-frequency regime,
we approximately rewrite the probability amplitudes as the leading order $a_j(t')=a_j^{(0)}(t')=A_j(t')$.
Thus, $|A_j|^2=|c_j|^2 \ (j=1,2,...,6)$ denote the probabilities of finding the two bosons in the six different Fock states in Eq. (2).
According to the perturbation analysis in the Appendix, we readily
obtain that such amplitudes are the slowly-varying functions
in time, which satisfy the following linear equations with constant
coefficients,
\begin{eqnarray}
&&i\frac{dA_1}{dt'}=2\epsilon^2(A_1\rho_1+A_2\rho_2),\nonumber\\
&&i\frac{dA_2}{dt'}=2\epsilon^2[2A_2\rho_1+(A_1+A_3)\rho_2],\nonumber\\
&&i\frac{dA_3}{dt'}=2\epsilon^2(A_3\rho_1+A_2\rho_2);\\ \label{eq6}
&&i\frac{dA_4}{dt'}=-\epsilon\mathcal{J}_{0}(\frac{\varepsilon}{\omega})A_5-2\epsilon^2(2A_4\rho_1+A_6\rho_2),\nonumber\\
&&i\frac{dA_5}{dt'}=-\epsilon\mathcal{J}_{0}(\frac{\varepsilon}{\omega})(A_4+A_6),\nonumber\\
&&i\frac{dA_6}{dt'}=-\epsilon\mathcal{J}_{0}(\frac{\varepsilon}{\omega})A_5-2\epsilon^2(2A_6\rho_1+A_4\rho_2),\label{eq7}
\end{eqnarray}
where $\rho_i$ ($\ i=1,2$) are set as (see Appendix)
\begin{equation}\label{eq8}
\rho_1=\sum\limits_{n'=-\infty}^{\infty}\frac{\mathcal{J}^2_{n'}(\frac{\varepsilon}{\omega})}{\frac{U_0}{\omega}+n'},~
\rho_2=\sum\limits_{n'=-\infty}^{\infty}\frac{\mathcal{J}_{n'}(\frac{\varepsilon}{\omega})\mathcal{J}_{-n'}(\frac{\varepsilon}{\omega})}{\frac{U_0}{\omega}+n'}
\end{equation}
for $U_0/\omega+n'\ne 0$. Therefore, Eqs. (6) and (7) are always definable and applicable except for the resonant case in which $\rho_i$ tends infinity. It is worth noting that for a stronger reduced interaction obeying $|u|>J$ at least, the value of any term in the summations of Eq. (8) is less than $\epsilon^{-1}$ such that $\epsilon^2\rho_i$ may be a second-order quantity and Eqs. (6) and (7) could be applicable as a set of second-order equations. In fact, $|\rho_i|<\epsilon^{-1}$ implies that the inequality $|U_0/\omega +n'|=|n+n'+u/\omega |>\epsilon=J/\omega$ holds for any pair $\{n\in [0,\ \infty),\ n'\in (-\infty,\ \infty)\}$, which results in $|u|>J$. Particularly, we will numerically prove that perfect applicability of the second-order perturbation method requires $|u|\ge 10 J$ later. Combining Eq. (6) with Eq. (7), we note that dynamics of the three paired states [the two bosons occupy the same site for $A_j(t')$ with $j=1,2,3$] is decoupled from that of the three unpaired states [the two bosons occupy distinct sites for $A_j(t')$ with $j=4,5,6$].

Clearly, for $|u|>J$, Eqs. (6) and (7) describe the second order approximation, where the time evolution of any paired state amplitude is a second order long-time-scale process, since its time derivative is proportional to only the second order constant $\epsilon^2$. The similar results have been seen previously in the tight-binding optical lattice \cite{Longhi2012two}.
The second-order coupling coefficients of Eqs. (6) and (7) are proportional to the parameter $\epsilon^2\rho_2$, which describes the second-order tunneling rate of the system. The nonzero tunneling coefficient means that the tunneling can occur, respectively, between the three paired states based on Eq. (6), and between the three unpaired states based on Eq. (7). In Fig. 2, we plot the factor $\rho_2$ of the second-order tunneling coefficients as a function of the driving parameters $\varepsilon/\omega$ and self-interaction intensity $U_0/\omega$, where Fig. 2(b) is the plan view of Fig. 2(a). Combining Eq. (8) with Fig. 2 we can see that the factor $\rho_2$ tends to infinity for any integer value of $U_0/\omega$ and arbitrary value ranges of $\varepsilon/\omega$, while its values are small enough for the considered far-resonant regime. Note that, in Fig. 2, the very great $\rho$ values are not shown, since we have avoided the integer values of $U_0/\omega$ through selecting a rational step such that the multiple-time-scale asymptotic analysis holds. In the second approximation, Eqs. (6) and (7) mean that the CDT between the paired states will occur provided that the condition $\rho_2=0$ is satisfied.
\begin{figure}[htp] \center
\includegraphics[width=2.2in]{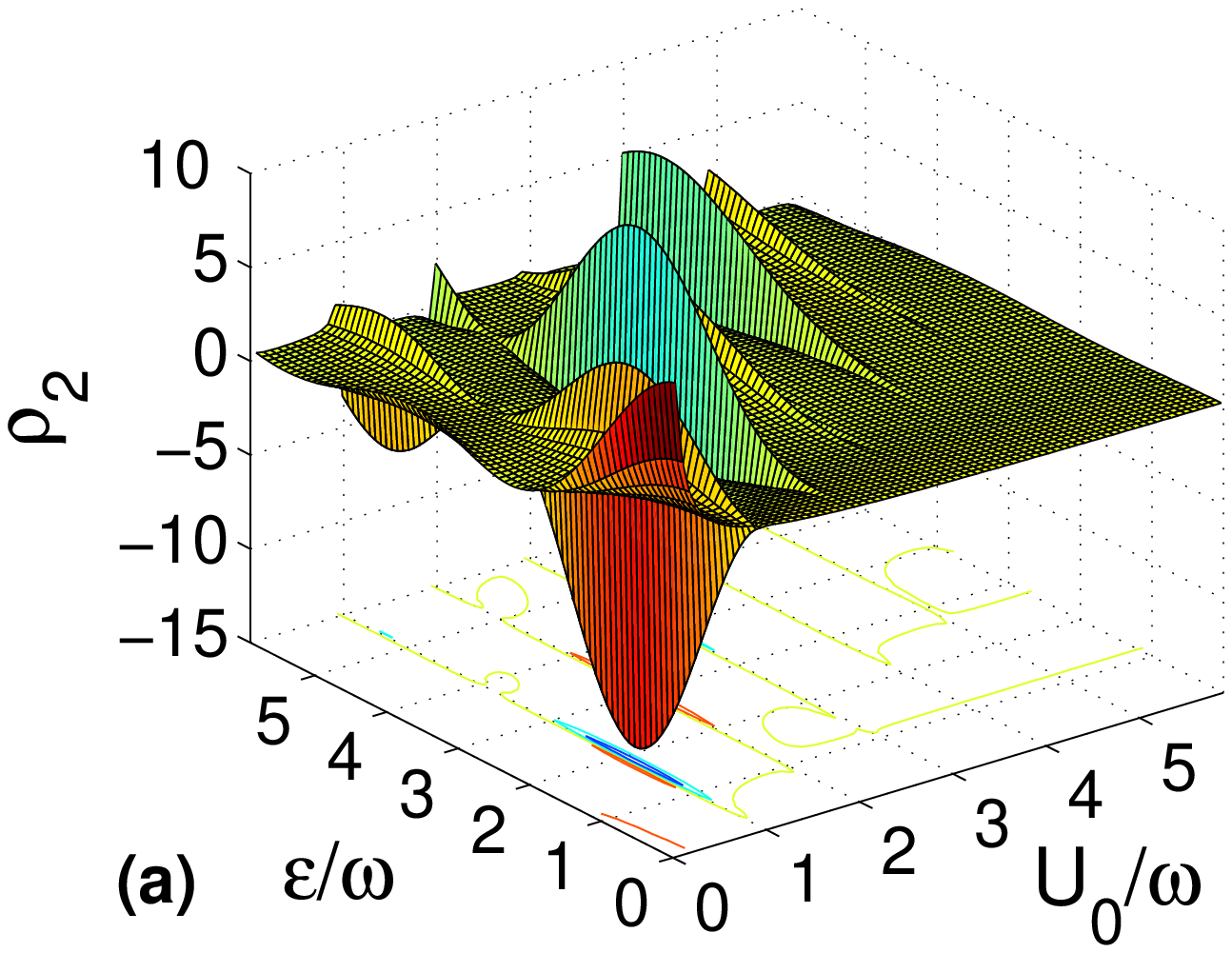}
\includegraphics[width=2.2in]{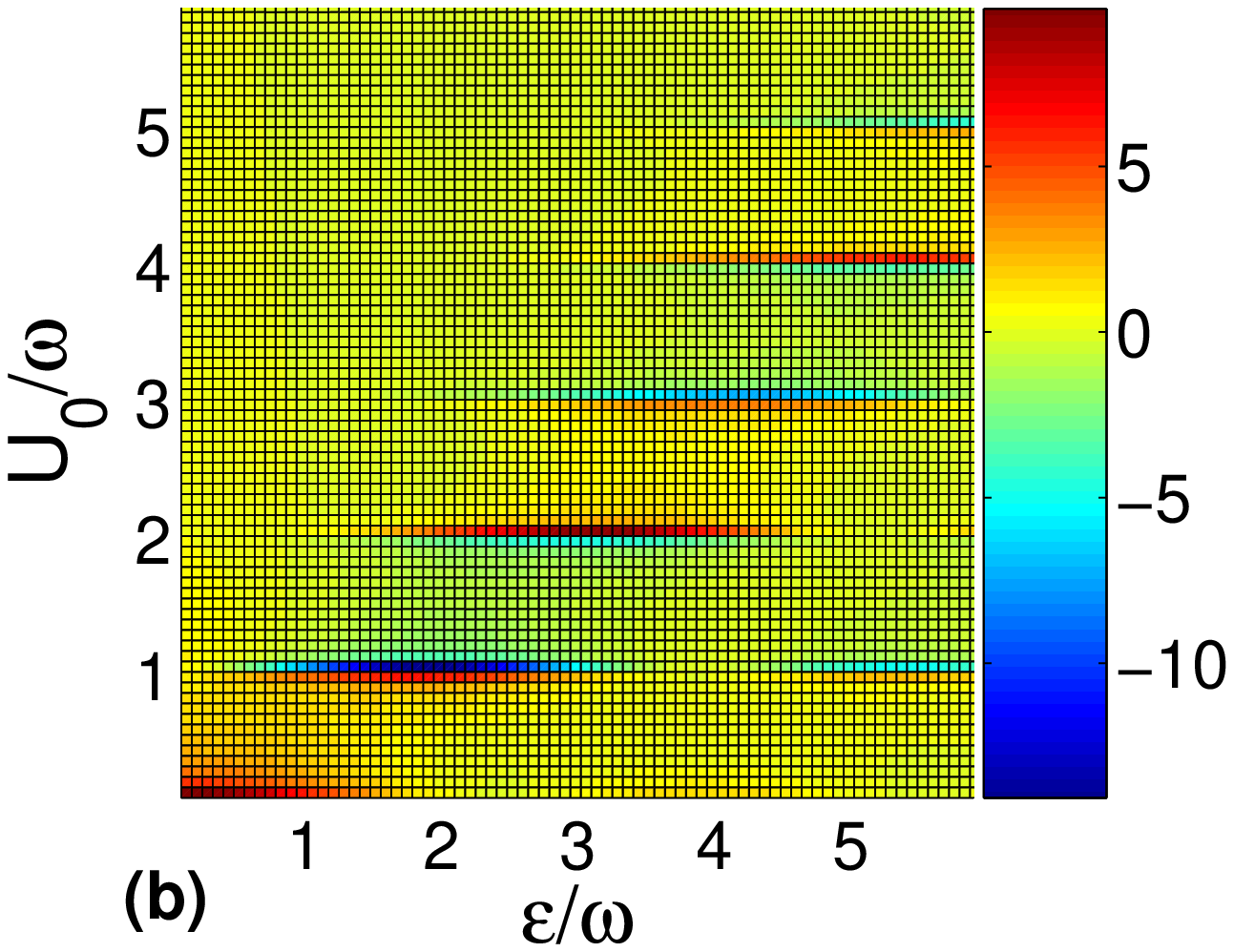}
\caption{\scriptsize{(Color online) The factor $\rho_2$ of the second-order tunneling coefficients as a function of $\varepsilon/\omega$ and $U_0/\omega$, defined by Eq. (8), where Fig. 2(b) is the plan view of Fig. 2(a). The infinite $\rho_2$ value at integer $U_0/\omega$ has been omitted.}}
\end{figure}

According to Eqs. (6) and (7), we make an exact comparison of tunneling rates between the three paired states and three unpaired states for two
different initial conditions as follows. Firstly, for the two bosons initially occupying the middle well [i.e., $P_2(0)=1,\ P_{j\ne 2}(0)=0$], we seek the analytical solutions
$P_j(t) (j=1,2,...,6)$ from Eqs. (6) and (7). To do so, we make the function transformations $A_1=A'_1\exp(-i\frac{2J^2\rho_1}{\omega}t)$, $A_2=A'_2\exp(-i\frac{4J^2\rho_1}{\omega}t)$ and
$A_3=A'_3\exp(-i\frac{2J^2\rho_1}{\omega}t)$. Inserting these expressions into Eq. (6) yields the coupled equations
\begin{eqnarray}
&&i\frac{dA'_1}{dt}=\frac{2J^2\rho_2}{\omega}A'_2e^{-i\frac{2J^2\rho_1}{\omega}t},\\ \label{eq9}
&&i\frac{dA'_2}{dt}=\frac{2J^2\rho_2}{\omega}e^{i\frac{2J^2\rho_1}{\omega}t}(A'_1+A'_3),\\ \label{eq10}
&&i\frac{dA'_3}{dt}=\frac{2J^2\rho_2}{\omega}A'_2e^{-i\frac{2J^2\rho_1}{\omega}t}.\label{eq11}
\end{eqnarray}
Combining Eq. (9) with Eq. (11) produces $\frac{d(A'_1+A'_3)}{dt}=\frac{4J^2\rho_2}{\omega}A'_2e^{-i\frac{2J^2\rho_1}{\omega}t}$. Eliminating $(A'_1+A'_3)$ from this equation and Eq. (10) yields the decoupled equation
\begin{equation}\label{eq12}
i\ddot{A}'_2+\frac{2J^2\rho_1}{\omega}\dot{A}'_2+i\frac{8J^4\rho_2^2}{\omega^2}A'_2=0.
\end{equation}
This is a second-order linear equation with constant
coefficients, whose general solution is well-known, $A'_2=A'_+\exp(\chi_+t)+A'_-\exp(\chi_-t)$ for the parameters $\chi_{\pm}=i(\frac{J^2}{\omega}\rho_1\pm\frac{J^2}{\omega}\sqrt{\rho_1^2+8\rho_2^2})$ and the undetermined constants $A'_{\pm}$ adjusted by the initial conditions. Under the above initial conditions the general solution becomes the special one
\begin{equation}\label{eq13}
A'_2(t)=e^{i\frac{J^2\rho_1}{\omega}t}\Big[\cos(\omega_1t)-i\frac{\rho_1}{\rho_1^2+8\rho_2^2}\sin(\omega_1t)\Big],
\end{equation}
with $\omega_1=\frac{J^2}{\omega}\sqrt{\rho_1^2+8\rho_2^2}$. Thus under the initial conditions $P_2(0)=1,\ P_{j\ne 2}(0)=0$, the analytical probabilities of Eqs. (6) and (7) are constructed as $P_4(t)=P_5(t)=P_6(t)=0$,
\begin{equation}\label{eq14}
P_2(t)=|A'_2(t)|^2=\frac{\rho_1^2}{\rho_1^2+8\rho_2^2}+\frac{8\rho_2^2}{\rho_1^2+8\rho_2^2}\cos^2(\omega_1t),
\end{equation}
\begin{equation}\label{eq15}
P_1(t)=P_3(t)=\frac{4\rho_2^2}{\rho_1^2+8\rho_2^2}\sin^2(\omega_1t).
\end{equation}
\begin{figure}[htp] \center
\includegraphics[width=1.6in]{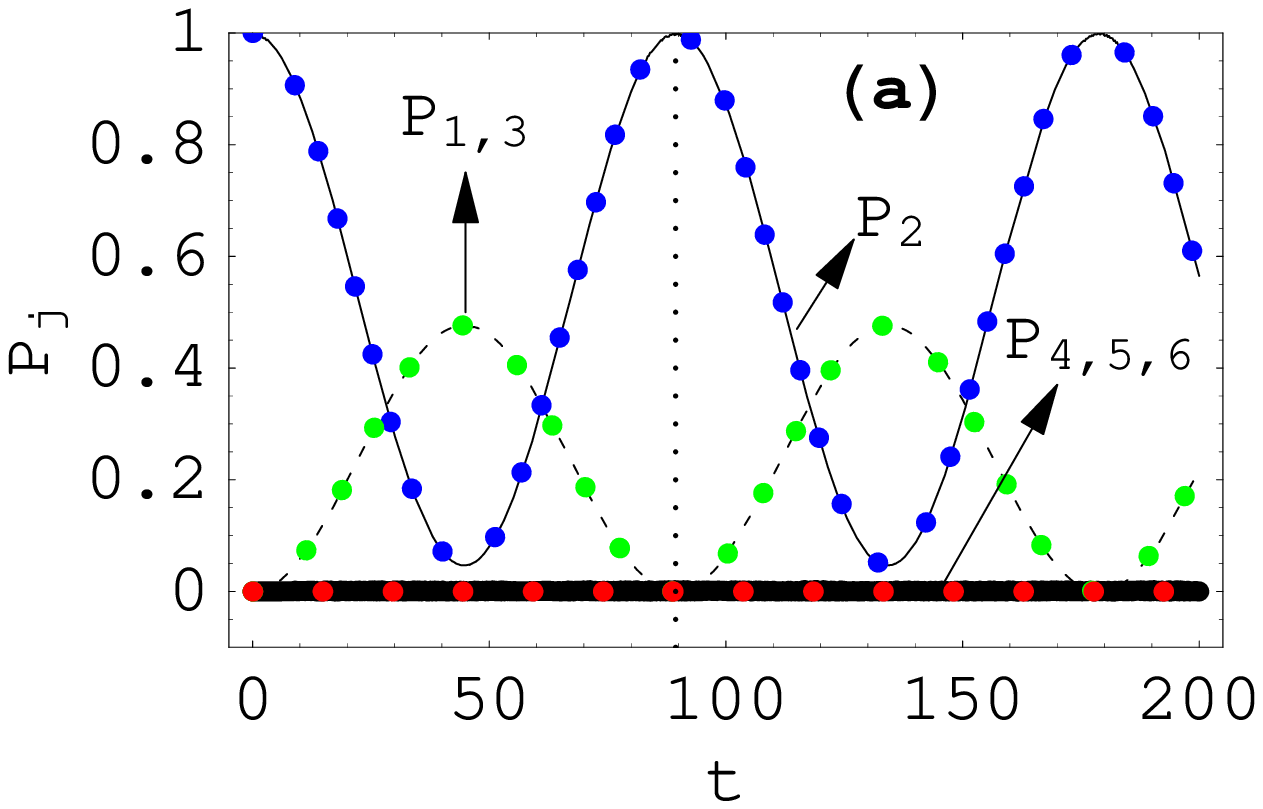}
\includegraphics[width=1.6in]{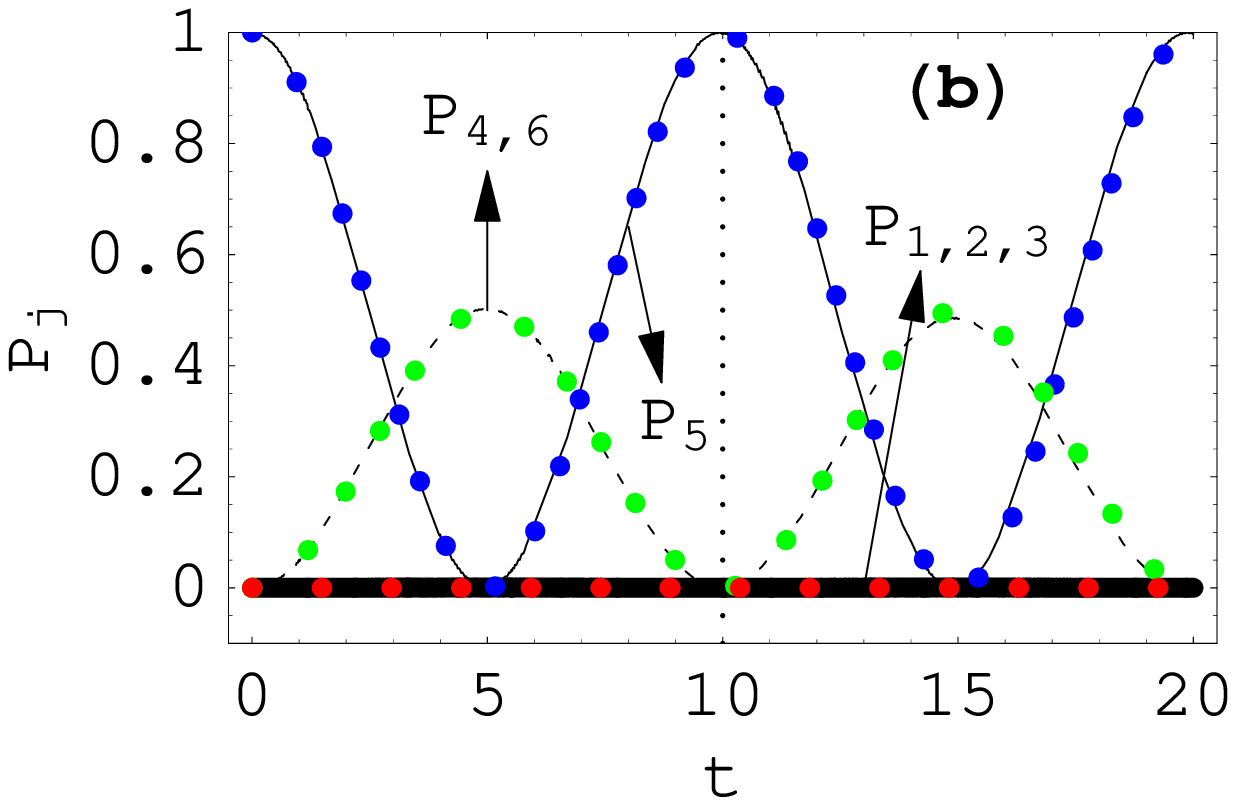}
\caption{\scriptsize{(Color online) Time evolutions of the probabilities $P_j$ ($j=1,2,...,6$) for the parameters $J=1$, $\omega=80$, $\varepsilon/\omega=2$, $U_0/\omega=1.5$ and the initial conditions (a) $P_2(0)=1,\ P_{j\ne 2}(0)=0$; (b) $P_5(0)=1,\ P_{j\ne 5}(0)=0$.
In (a), the dashed line corresponds to the probabilities $P_{1,3}$, and the
thin and the thick solid lines indicate the probabilities $P_2$ and $P_{4,5,6}$, respectively. In (b), the dashed line corresponds to the probabilities $P_{4,6}$, and the
thin and the thick solid lines indicate the probabilities $P_5$ and $P_{1,2,3}$, respectively. In this figure and the following figures, all the circular points indicate the analytical
solutions and the curves represent the numerical results.}}
\end{figure}
Secondly, for the two bosons initially occupying the left and middle well, respectively, [i.e., $P_5(0)=1,\ P_{j\ne 5}(0)=0$], we easily obtain the analytical solutions
$P_1(t)=P_2(t)=P_3(t)=0$, $P_5(t)=\cos^2(\omega_2t)$, and $P_4(t)=P_6(t)=\frac{1}{2}\sin^2(\omega_2t)$ with $\omega_2=\sqrt{2}J\mathcal{J}_{0}(\frac{\varepsilon}{\omega})$, if we neglect the second-order small quantities of Eq. (7) in the high-frequency regime. As an example, selecting the parameters $\omega=80\gg J=1$, $\varepsilon/\omega=2$ and $U_0/\omega=1.5$. Therefore, the tunneling period $T_1=\pi/\omega_1\approx 89$ corresponding to the second-order tunneling effect, and the tunneling period $T_2=\pi/\omega_2\approx 10$ corresponding to the first-order
tunneling effect.

In Fig. 3, we numerically plot the time evolutions of the probabilities $P_j$ ($j=1,2,...,6$) based on Eq. (3) for the above two
initial conditions, and the circular points correspond to the above analytical results. Obviously, the analytical results are in perfect agreement with the numerical simulations. The zero probability of the unpaired states in Fig. 3(a) and the zero probability of the paired states in Fig. 3(b) mean the selected CDT between the paired states and unpaired states.
\begin{figure}[htp] \center
\includegraphics[width=2.2in]{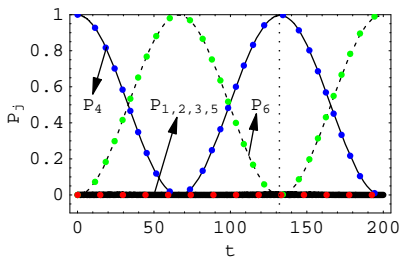}
\caption{\scriptsize{(Color online) Time evolutions of the probabilities $P_j$ ($j=1,2,...,6$) for the initial conditions $P_4(0)=1,\ P_{j\ne 4}(0)=0$. The driving parameter is set as $\varepsilon/\omega=2.405$, and the other parameters are the same as those of
Fig. 3(b).}}
\end{figure}

In the high-frequency regime, we note that the second-order correction term is much less than the first-order
term, as shown in Eq. (7), and the dominating dynamics is decided by the first-order correction term provided that it does not vanish, on the contrary, is decided by the high-order correction term if the first-order correction term vanishes in the case of $\mathcal{J}_{0}(\frac{\varepsilon}{\omega})=0$. As another example, in Fig. 4, we numerically show the time evolutions of the probabilities $P_j$ ($j=1,2,...,6$) based on Eq. (3) for two bosons initially occupying the left and middle well, respectively [i.e., $P_4(0)=1,\ P_{j\ne 4}(0)=0$], where the parameter is set as $\varepsilon/\omega=2.405$ corresponding to the first zero of $\mathcal{J}_{0}(\frac{\varepsilon}{\omega})$ and the other parameters are the same as those of Fig. 3(b). In this case, we readily calculate the probabilities analytically, because of $\dot {A}_5(t)=0$ in Eq. (7). Under the initial conditions $A_4(0)=1$ and $A_{j\ne 4}(0)=0$, we immediately obtain  the analytical solutions $A_1=A_2=A_3=A_5=0,\ A_4=\exp(i\frac{4J^2\rho_1}{\omega}t)\cos(\frac{2J^2\rho_2}{\omega}t)$ and $A_6=\exp(i\frac{4J^2\rho_1}{\omega}t)\sin(\frac{2J^2\rho_2}{\omega}t)$. Thus, the corresponding probabilities of Eqs. (6) and (7) read
$P_1=P_2=P_3=P_5=0$, $P_4=|A_4(t)|^2=\cos^2(\frac{2J^2\rho_2}{\omega}t)$ and $P_6=\sin^2(\frac{2J^2\rho_2}{\omega}t)$, which are plotted by the circular points in Fig. 4.
The analytical and numerical results consistently verify that the tunneling period in Fig. 4 is about $130$, which is in the same order of magnitude as the above $T_1$ for a second order tunneling period.

\begin{figure}[htp] \center
\includegraphics[width=2.2in]{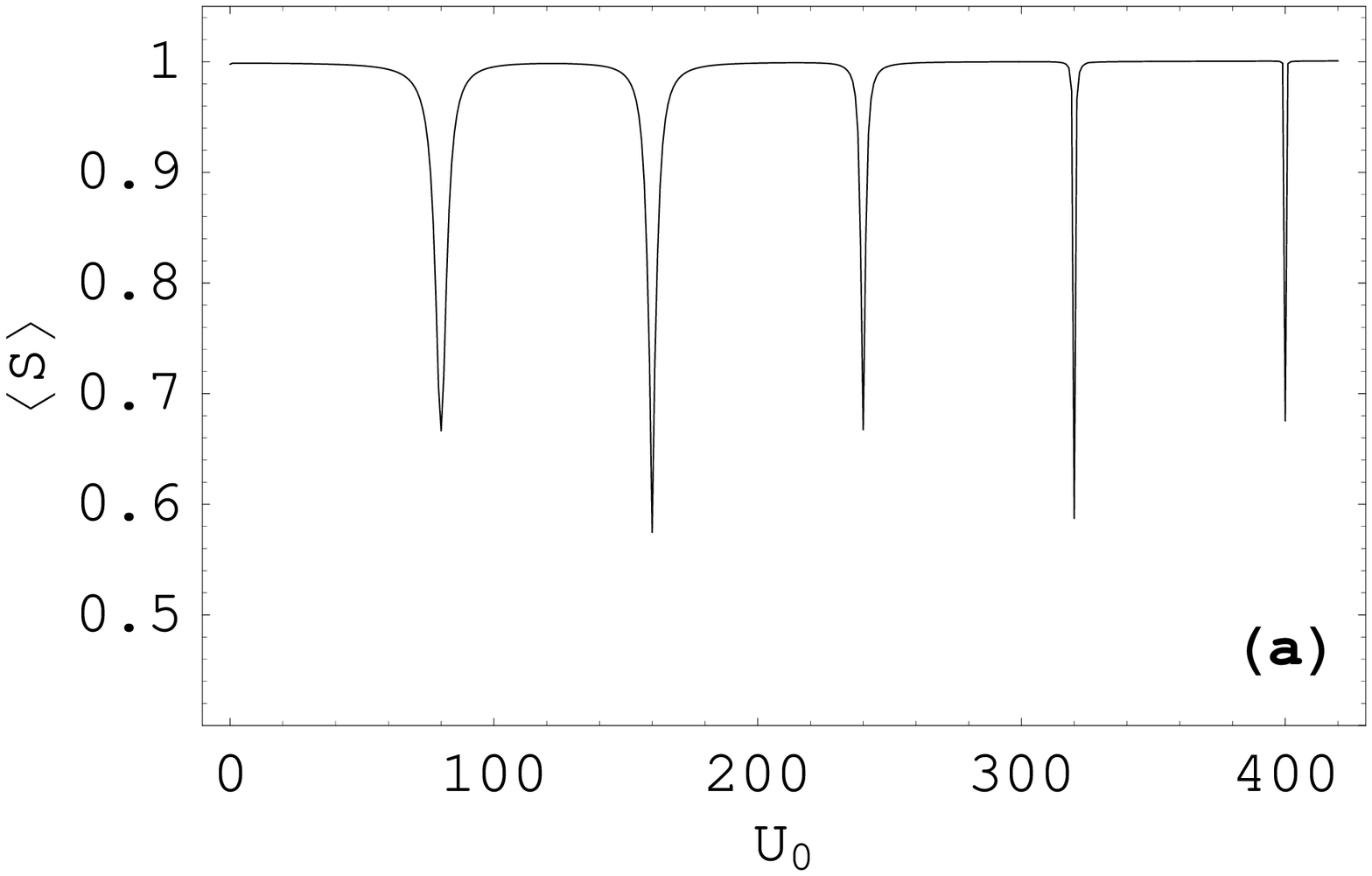}
\includegraphics[width=2.2in]{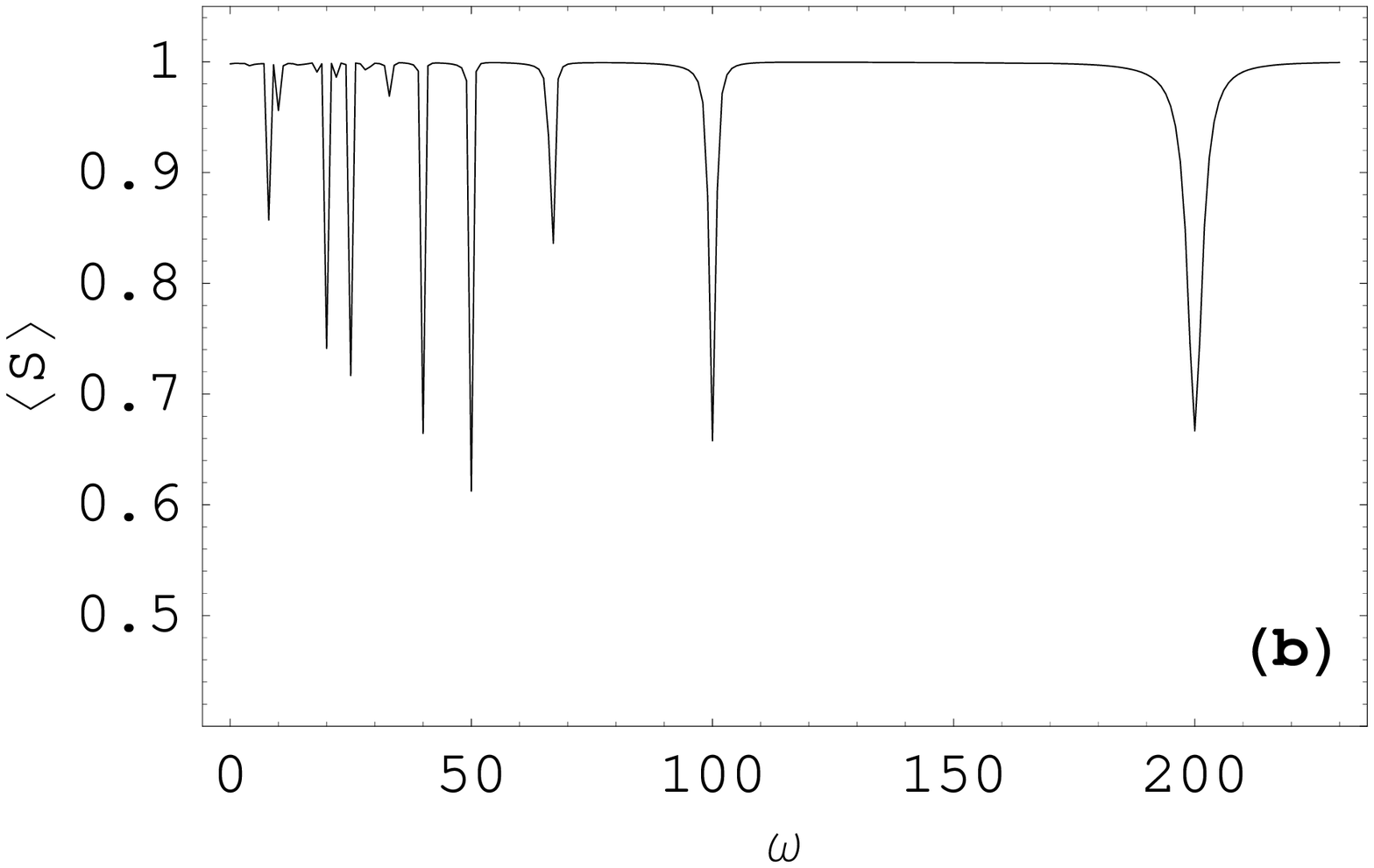}
\caption{\scriptsize{The time-averaged total probability $\langle S\rangle$ versus the self-interaction $U_0$ in (a) and versus the driving frequency $\omega$ in (b), computed numerically based on Eq. (3). (a) the parameters are the same as those of Fig. 1 except for $U_0$, and (b) $J=1$, $\varepsilon=160$, and $U_0=200$. The time used for averaging is 200 in dimensionless units.}}
\end{figure}

In order to further confirm the analytical results in Eqs. (4), (6) and (7), we define the time-averaged total probability of finding the two interacting bosons in the three paired states as $\langle S\rangle=\langle P_1\rangle+\langle P_2\rangle+\langle P_3\rangle=\frac {1}{ \tau}\int_0^{\tau}(P_1+P_2+P_3)dt$ for $\tau=200J$. The normalization means the time-averaged total probability in the three unpaired states being $1-\langle S\rangle$. Taking the initial conditions $P_2(0)=1, P_{j\ne 2}(0)=0$ and parameter $J=1$, from Eq. (3) we numerically give $\langle S\rangle$ as the function of the self-interaction $U_0$ for $\omega=80$ and $\varepsilon/\omega=
2.405$, as in Fig. 5(a). It is shown that the time-averaged total probability in the three paired states possesses different features, for the multiphoton resonant points, the far-resonant regions and the near-resonant regions, respectively. Firstly, at each of the resonant points, i.e., $U_0=m\omega$ with $m=1,2,...,5$ being integer, $\langle S\rangle$ drops to the lowest points which mean that the separation probability $1-\langle S\rangle$ of the two bosons is the largest, as shown in Fig. 1. Secondly, the two bosons can also be separated in the near-resonant regions, however, the time-averaged total probability $\langle S\rangle$ tends to one and the separation probability tends to zero rapidly as increasing the reduced interaction strength $|u|$. Finally, in the far-resonant regions with larger $|u|$ values, $\langle S\rangle$ is always equal to 1. Let half-width of the valley centred at $m$th resonant point of Fig. 5(a) be $|u|_m$. The largest half-width for fitting $\langle S\rangle\approx 1$ can be estimated as $|u|_1 \approx 10 J$ from the first valley. This indicates that a selected CDT between the paired states and the unpaired states can happen in the region $10 J\le |u|\le \omega/2$, which is called the \emph{far-resonant strongly-interacting regime} in this paper. Such a CDT enables the two bosons form a stable bound pair and
cannot move independently for a stronger reduced interaction. Similar to Refs. \cite{Winkler, Folling},
the phenomena can be understood that potential energy of two bosons occupying a single well for strong repulsive interaction is greater than the maximum
kinetic energy of two separate bosons, according to the principle of conservation of energy, the two bosons only forming a
stable bound pair tunnel from a well to a neighboring well in the triple-well without dissipation. Only for the resonant case,
the boson pair can be separated, because the bosons could absorb photons from the ac driving field.
In the weakly-interacting regime, CDT is expected to occur in Fig. 5(a) because $\varepsilon/\omega=2.405$
is the first root of $\mathcal{J}_{0}(x)=0$, and we will further consider it in the next section. To show that the above analysis
is generic in the high-frequency regime, we plot the time-averaged total probability $\langle S\rangle$ as a function of the
driving frequency $\omega$ in Fig. 5(b) with $\varepsilon=160$ and
$U_0=200$. Fig. 5(b) explicitly shows that the lowest points of $\langle S\rangle$ appear at the  resonant points $U_0/\omega=5,4,3,2,1$ for the sufficiently high frequency, $\omega> 30$. The results agree with the above analysis on Fig. 5(a).

It is well known that CDT can occur at the collapse points of the Floquet quasienergy spectrum \cite{Longhi2008, Creffield2009},
so the above-mentioned tunneling properties will be confirmed by the Floquet quasienergy analysis as follows.

\begin{figure}[htp] \center
\includegraphics[width=2.2in]{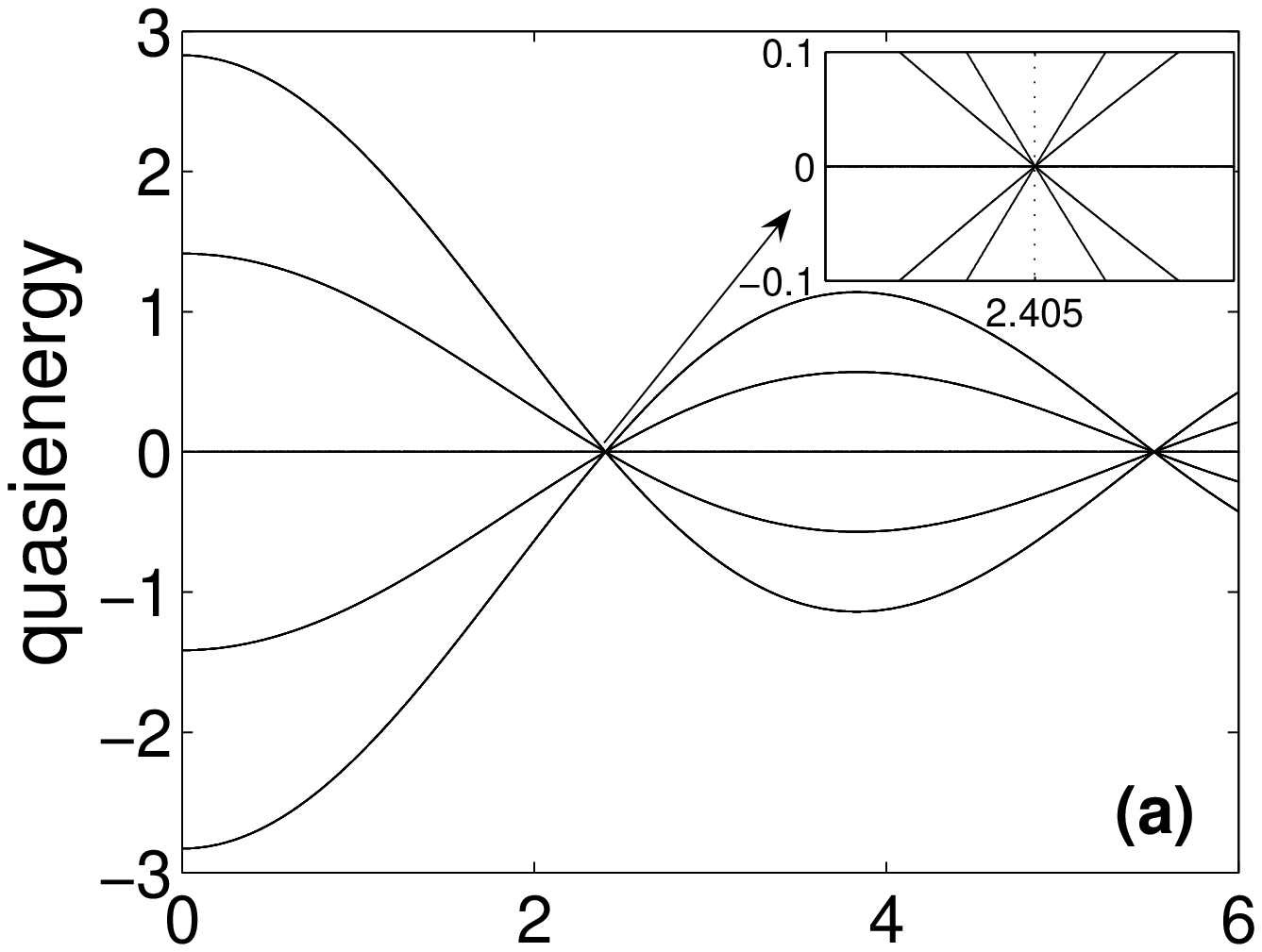}
\includegraphics[width=2.2in]{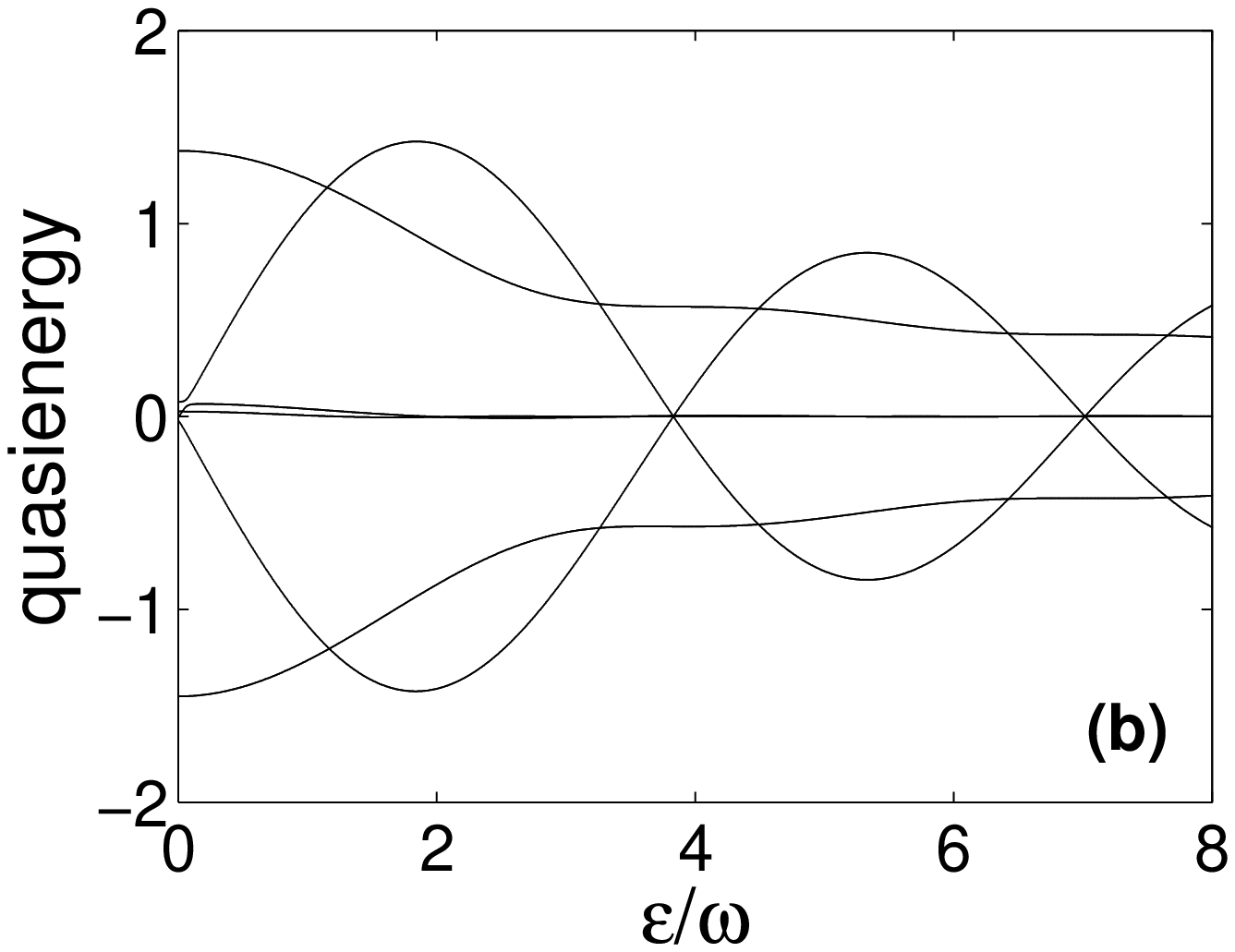}
\caption{\scriptsize{Numerical quasienergy spectrum versus $\varepsilon/\omega$ for the self-interaction $U_0=m\omega, \ m=0,1$, respectively. The parameters are set as $J=1$,
$\omega=80$, and (a) $U_0=0$, (b) $U_0=80$. The inset shows an enlargement of quasienergies near the point $\varepsilon/\omega=2.405$ corresponding to the first zero of $\mathcal{J}_{0}(\varepsilon/\omega)$.}}
\end{figure}

\section{Floquet quasienergy analysis}

The Floquet theory provides a powerful tool to analyze the dynamics of a time-periodic quantum system \cite{Shirley}. According to the Floquet
theory, the solutions of the time-dependent Schr\"{o}dinger equation can be written as $|\psi_k(t)\rangle=e^{-i E_kt}|\phi_k(t)\rangle$,
with $|\phi_k(t)\rangle$ being the Floquet states and $E_k$ Floquet quasienergies. In analogy to the Bloch solutions for the spatially
periodic system, the quasienergy can only be determined up to a integer multiple of the photon energy $\omega$, and for the sake of definiteness
it is usually assumed to vary in the first Brillouin zone $-\omega/2<E\leq\omega/2$. The Floquet states inherit the period of the Hamiltonian,
and are eigenstates of the time evolution operator for one period of the driving
\begin{equation}\label{eq16}
U(T,0)=\mathcal{T}\exp\Big[-i\int^T_0H(t)dt\Big],
\end{equation}
where $\mathcal{T}$ is the time-ordering operator and $T=2\pi/\omega$ is the period of the driving. Noticing that eigenvalues of $U(T,0)$ are
$\exp(-iE_kT)$, the quasienergies of this system can be determined directly so long as we diagonalize $U(T,0)$. In Fig. 6, selecting the parameters as $J=1$ and $\omega=80$, we
show the numerical results of the quasienergy spectra as the functions of driving parameters $\varepsilon/\omega$ for $U_0=m\omega, \ m=0,1$ with zero reduced interaction, respectively.
In Fig. 6(a), for two noninteracting bosons, the quasienergy spectrum shows collapses at some fixed values of the driving parameters for which $\mathcal{J}_{0}
(\frac{\varepsilon}{\omega})=0$. The inset of Fig. 6(a) is an enlargement of quasienergies near the collapse point $\varepsilon/\omega
=2.405$, corresponding to the first zero of $\mathcal{J}_{0}(\frac{\varepsilon}{\omega})$ and shows an exact level-crossing at $\varepsilon/\omega\approx 2.405$,
analogous to a single boson in a triple-well system \cite{Gengbiao}. In the resonant regime with $U_0=\omega=80$, the quasienergies of Fig. 6(b) show that the crossings of some quasienergies and the avoided crossing of the other quasienergies appear at the zero points $(\varepsilon/\omega=3.832, ...)$ of the first-order Bessel function $\mathcal{J}_{1}(\varepsilon/\omega)$. The numerical results are in good
agreement with the analytical results from Eq. (4) with $m=1,\ u=0$ in second section. In the case $u=0$, Eqs. (6) and (7) are no longer valid and any quasienergy may be associated with both the paired states and the unpaired states.

\begin{figure}[htp] \center
\includegraphics[width=2.2in]{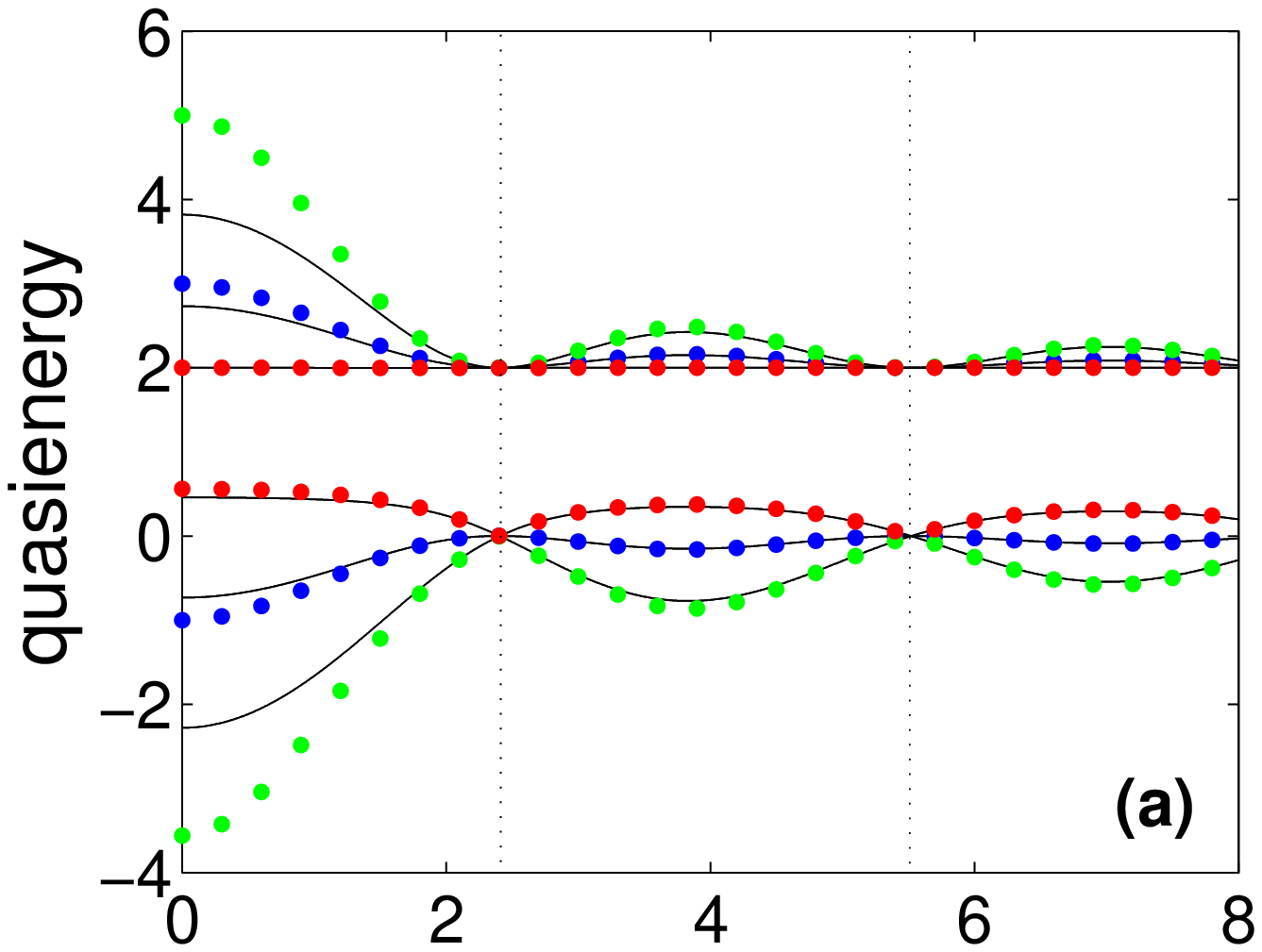}
\includegraphics[width=2.2in]{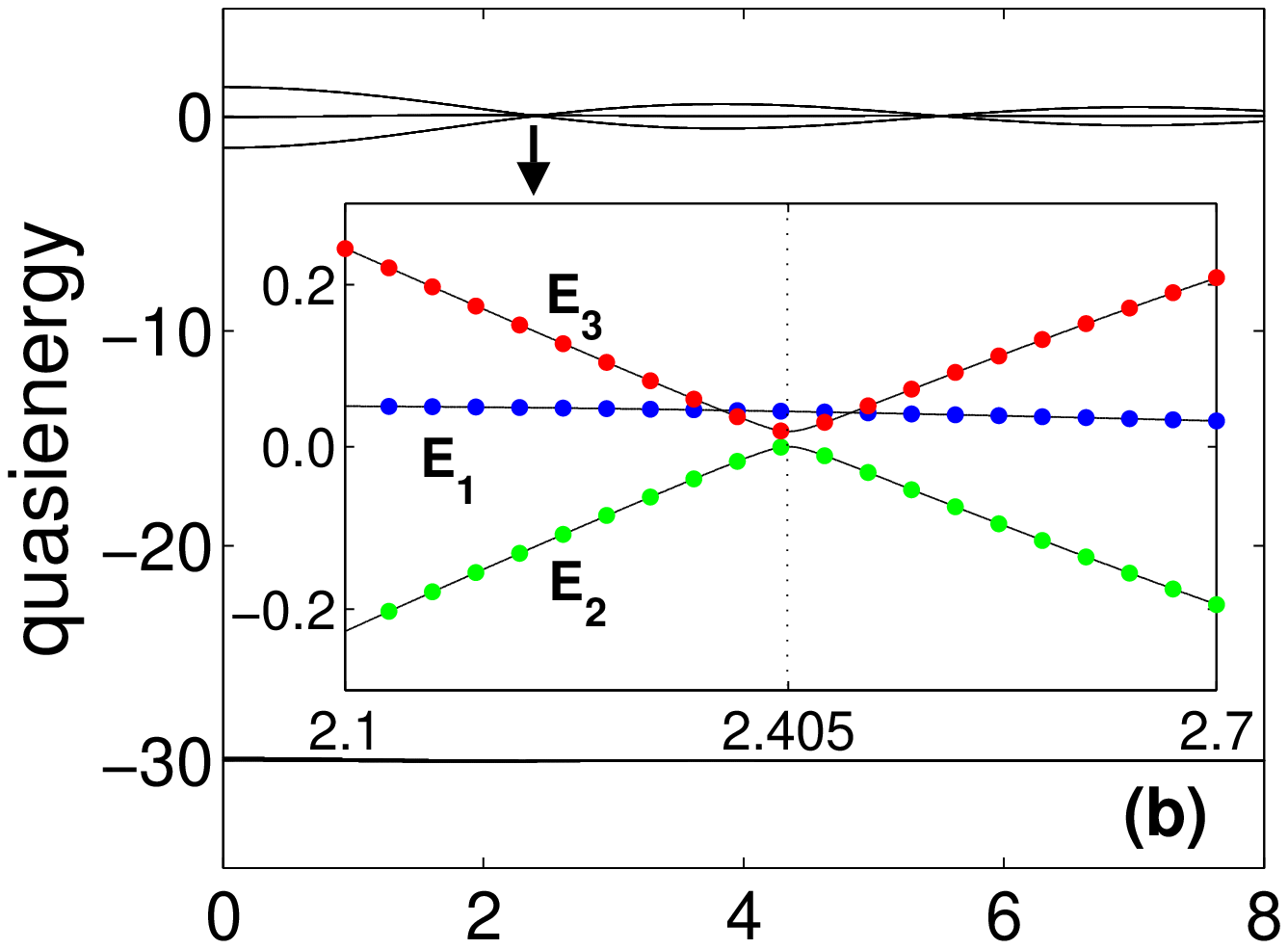}
\caption{\scriptsize{(Color online) Numerical quasienergy spectrum versus $\varepsilon/\omega$ for (a) $U_0=u=2J=2$ and (b) $U_0=\omega-30=50,\ u=-30$. The other parameters are the same as those of Fig. 6. In Fig. 7(a), the above three curves denote the quasienergies of paired states and the below ones the quasienergies of unpaired states, while the situation is contrary to Fig. 7(b). The circular points are associated with the perturbation results from Eqs. (19), (20) and (23), the thin dotted lines indicate the zero points of $\mathcal{J}_{0}(\varepsilon/\omega)$ and the arrow in Fig. 7(b) indicates the amplification position of the inset.}}
\end{figure}

When the reduced interaction strength is sufficiently larger (e.g. $|u|>J$), the quasienergy spectrum is divided into two energy bands which correspond to the three paired states and the three unpaired states, respectively, as shown in Fig. 7, where the quasienergies of paired states (or unpaired states) aperiodically oscillate near $u$ values (or $0$ value), so width of the energy gap between the two bands is proportional to the $|u|$ value.
For a weaker interaction with $U_0=u=2$, CDT between the different paired states and between the different unpaired states can be realized for the same driving parameters, as indicated by the level-crossing points in Fig. 7(a),
when the ratio of the field amplitude $\varepsilon$ and the field frequency $\omega$ is a root of the equation $\mathcal{J}_{0}(\varepsilon/\omega)=0$. Precise agreements between the numerical results based on Eq. (3) and the analytical results from Eqs. (19), (20) and (23) are observed in Fig. 7(a) for a sufficiently larger range of ratio $\varepsilon/\omega$. The small deviation between the both results in Fig. 7(a) indicates that the second-order perturbation method is perfectly applicable only for some suitable parameter regions. We have also investigated the quasienergy spectra for $|u|=5,6,...$ which are not shown here. \emph{The results conformably verify that in the far-resonant strongly-interacting regime, $\omega/2 \ge |u|\ge |u|_1\approx 10 J$ is just the above suitable parameter regions}. Interestingly, the energy band corresponding to the paired states becomes narrower and the energy gap tends to wider as the increase of self-interaction intensity from $|u|=2<|u|_1$ to $|u|=30>|u|_1$. The wider gap means quantum transition between the both kinds of states is hard to occur, and the narrower band necessitates to analyze the Floquet quasienergy spectrum from both cases of the unpaired states and the paired states, respectively.

\subsection{Avoided level-crossing of unpaired states}

In the far-resonant strongly-interacting regime, selecting the parameters as $J=1$, $\omega=80$ and $U_0=50$, i.e., $m=1$, $u=-30$, from Eq. (3) we numerically plot quasienergy spectrum versus $\varepsilon/\omega$ in Fig. 7(b). In this figure, the quasienergies corresponding to the unpaired states shows collapses when $\varepsilon/\omega$
are the roots of $\mathcal{J}_{0}(\varepsilon/\omega)=0$, however, the energy band corresponding to the paired states has collapsed into an approximate straight line.
The inset of Fig. 7(b) is an enlargement of quasienergies corresponding to the three unpaired states near the first collapse point
$\varepsilon/\omega\approx 2.405$, and the fine structure of energy spectrum exhibits that the pseudocollapse point is converted to an avoided crossing point at
$\varepsilon/\omega\approx 2.405$ and two different crossing points due to the second-order correction terms in Eq. (7).

To explain the numerical result, from Eq. (7) we analytically calculate the quasienergies corresponding to the three unpaired states. Note that the period of functions
$\exp[-i\varphi(t)]$ and $\exp[\pm2i\varphi(t)]$ is $T$. Therefore, we can construct the Floquet states by setting \cite{Gengbiao}
$A_j(t)=B_j\exp(-iEt)$ ($j=4,5,6$) for the
three unpaired states with constant $B_j$, then rewriting Eq. (7) as the time-independent form
\begin{eqnarray}\label{eq17}
&&EB_4=-J\mathcal{J}_{0}(\frac{\varepsilon}{\omega})B_5-2\frac{J^2}{\omega}(2B_4\rho_1+B_6\rho_2),\nonumber\\
&&EB_5=-J\mathcal{J}_{0}(\frac{\varepsilon}{\omega})(B_4+B_6),\nonumber\\
&&EB_6=-J\mathcal{J}_{0}(\frac{\varepsilon}{\omega})B_5-2\frac{J^2}{\omega}(2B_6\rho_1+B_4\rho_2).
\end{eqnarray}
The existence condition for the non-trivial solution of Eq. (17) reads
\begin{equation}\label{eq18}
\left|
\begin{array}{clr}
E+4\frac{J^2}{\omega}\rho_1 & J{J}_{0}(\frac{\varepsilon}{\omega})& 2\frac{J^2}{\omega}\rho_2 \\
J{J}_{0}(\frac{\varepsilon}{\omega}) & E & J{J}_{0}(\frac{\varepsilon}{\omega}) \\
2\frac{J^2}{\omega}\rho_2 & J{J}_{0}(\frac{\varepsilon}{\omega}) & E+4\frac{J^2}{\omega}\rho_1
\end{array}
\right|=0.
\end{equation}
From Eq. (18) we obtain three Floquet quasienergies corresponding to the three unpaired states
\begin{eqnarray}\label{eq19}
&&E_1=-\frac{2(2J^2\rho_1-J^2\rho_2)}{\omega},\nonumber\\
&&E_2=-\frac{-2J^2\rho_1-J^2\rho_2-\rho_3}{\omega},\nonumber\\
&&E_3=-\frac{-2J^2\rho_1-J^2\rho_2+\rho_3}{\omega},
\end{eqnarray}
where we have set
\begin{equation}\label{eq20}
\rho_3=\sqrt{(2J^2\rho_1+J^2\rho_2)^2+2J^2\omega^2\mathcal{J}_{0}^2(\frac{\varepsilon}{\omega})}.
\end{equation}
We now compare the analytical results of Eqs. (19) and (20) with the numerical computation based on the original Eq. (3). A
typical behavior of quasienegies near the first crossing point is plotted in the inset of Fig. 7(b). It is clearly
shown that the analytical result (the circular points) is in perfect agreement with the direct numerical
computation (the curves).

\subsection{A fine structure of quasienergy spectrum of paired states}

Next, we examine some detailed features of the quasienergies corresponding to the three paired states. In Ref. \cite{Longhi2012two}, Longhi et al proposed that in a lattice system, CDT can be realized between the paired states and between the unpaired states for the same parameters, namely the field parameters take the second root $\varepsilon/\omega=5.52$ of
$\mathcal{J}_{0}(\varepsilon/\omega)=0$ and the interaction intensity obeys $U_0/\omega=2.58$ corresponding to $\rho_2=0$. Here for the triple-well system we prove the similar result, and exhibit a fine structure of quasienergy spectrum of paired states, based on analytical Floquet solutions of Eqs. (6), (7) and (8).
According to Eq. (6), CDT occurs between the paired states if the condition $\rho_2=0$ is satisfied. From Eq. (8), we have $\rho_2=0$
at $\varepsilon/\omega\approx0.95$ in the region $\varepsilon/\omega\in[0,8]$ for $U_0/\omega=1.6$, and have $\rho_2=0$ at $\varepsilon/\omega\approx1.20, 2.02, 5.52, 5.74$ in the same region for $U_0/\omega=2.58$.
Selecting two different values of the self-interaction intensity, from Eq. (3) we numerically plot quasienergy spectrum versus $\varepsilon/\omega$ in Figs. 8(a) and 8(b), respectively,
with the insets being enlargements of quasienergies of the three paired states. At the points fitting $\rho_2=0$,
the level-crossing of two quasienergies will occur, and this indicates that CDT for paired states can
be observed at the crossing points of the partial levels. The predictions of the perturbation analysis can be confirmed by
direct numerical computation of the temporal evolution of the boson occupation probabilities from Eq. (3) (not depicted here).
\begin{figure}[htp] \center
\includegraphics[width=2.2in]{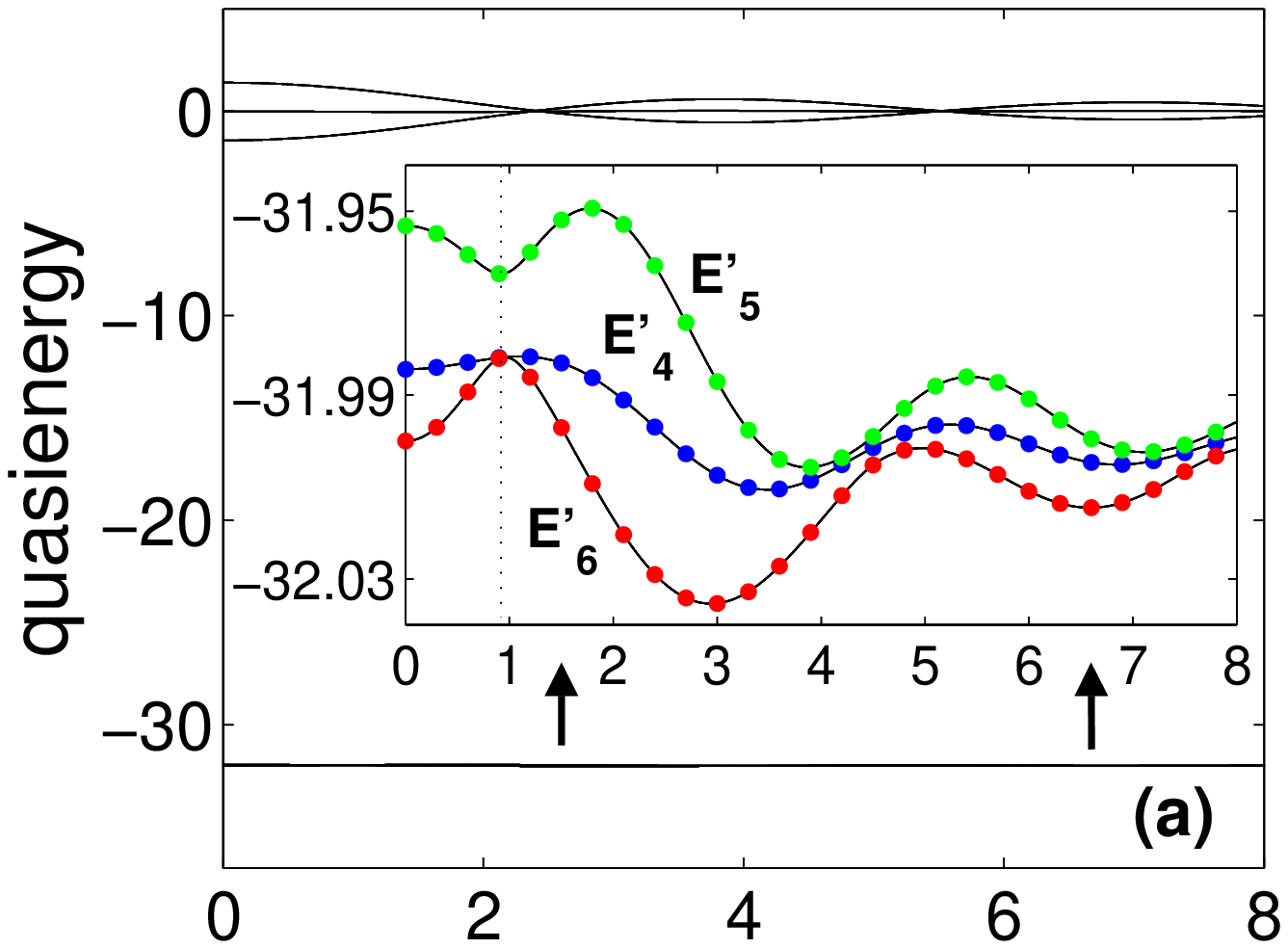}
\includegraphics[width=2.2in]{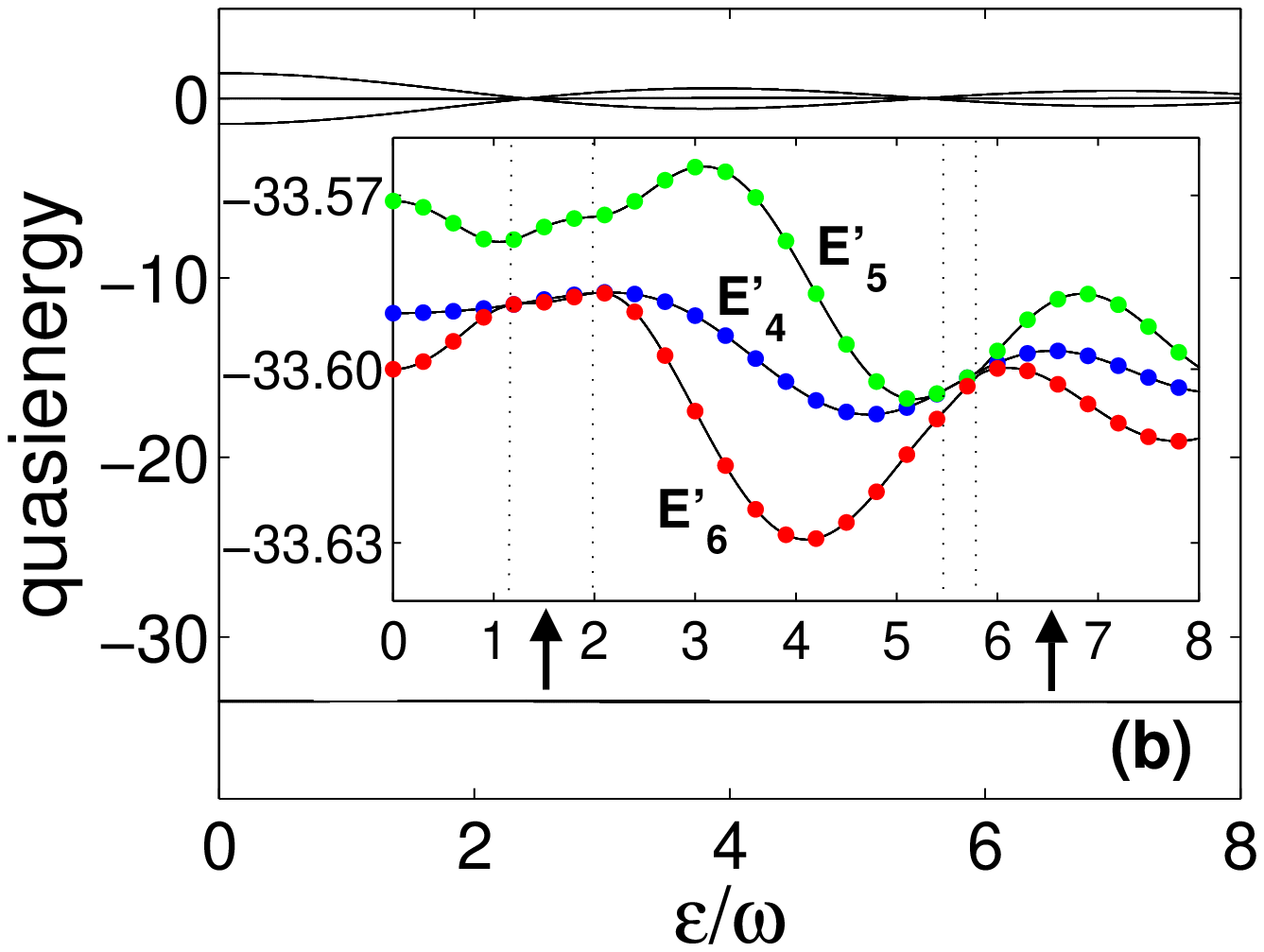}
\caption{\scriptsize{(Color online) Numerical results of quasienergy spectrum versus $\varepsilon/\omega$ for the stronger reduced interaction case. (a) $U_0/\omega=1.6$, (b) $U_0/\omega=2.58$. The other parameters are the same as those of Fig. 5. The arrows in each figure label the quasienergies of the paired states, and indicate the corresponding amplifications in the insets. The circular points in the insets are associated with the perturbation result from Eq. (23), and the thin dotted lines indicate the zero points of $\rho_2$.}}
\end{figure}

Following, we analytically calculate the quasienergies corresponding to the three paired states. Substituting the Floquet solutions
$A_j(t)=B_j\exp[-i(E-U_0)t]$ ($j=1,2,3$) into Eq. (6), we obtain easily
\begin{eqnarray}\label{eq21}
&&(E-U_0)B_1=2\frac{J^2}{\omega}(B_1\rho_1+B_2\rho_2),\nonumber\\
&&(E-U_0)B_2=2\frac{J^2}{\omega}[2B_2\rho_1+(B_1+B_3)\rho_2],\nonumber\\
&&(E-U_0)B_3=2\frac{J^2}{\omega}(B_3\rho_1+B_2\rho_2)
\end{eqnarray}
with the existence condition of the non-trivial solution
\begin{eqnarray}\label{eq22}
&& \Big(E-U_0-4\frac{J^2}{\omega}\rho_1\Big)\Big(E-U_0-2\frac{J^2}{\omega}\rho_1\Big)^2\nonumber\\ &&-8\frac{J^4}{\omega^2}\rho_2^2\Big(E-U_0-2\frac{J^2}{\omega}\rho_1\Big)=0.
\end{eqnarray}
From this equation we obtain three Floquet quasienergies corresponding to the three paired states as
\begin{eqnarray}\label{eq23}
&&E_4=U_0+2\frac{J^2}{\omega}\rho_1,\nonumber\\
&&E_5=U_0+\frac{3J^2\rho_1+\sqrt{J^4\rho_1^2+8J^4\rho_2^2}}{\omega},\nonumber\\
&&E_6=U_0+\frac{3J^2\rho_1-\sqrt{J^4\rho_1^2+8J^4\rho_2^2}}{\omega}.
\end{eqnarray}
We note that the quasienergies $E_j$ for $j=4,5,6$ should be converted to $E'_j$ in the first Brillouin zone, for example,
$U_0=2.58\omega\approx 206.4$, so the quasienergies are rewritten as $E'_j=E_j-3\omega$ for $\omega=80$ ($j=4,5,6$). We plot the analytical quasienergies versus
$\varepsilon/\omega$ for $U_0/\omega=1.6$ and $2.58$ in the insets of Figs. 8(a) and 8(b) as the circular points, respectively, which are in perfect agreement
with the direct numerical computations (the curves) based on the original Eq. (3).

\section{Conclusion}

We have investigated the tunneling dynamics of two bosons in a high-frequency driven triple
well for a continuously-increasing interaction intensity, by means of the multiple-time-scale asymptotic analysis. In the obtained far-resonant strongly-interacting regime, we consider the second-order perturbed correction and show that the dominant tunneling effect of paired states is a second order process,
similar to two bosons in a driven optical lattice \cite{Longhi2012two}. For a stronger reduced interaction, we make an exact comparison of tunneling rates between the paired states and unpaired states, and find that two bosons initially occupying the same well would form a stable bound pair. The selected CDT between the paired states and the unpaired
states can occur for different values of interaction intensity.
However, for the near-resonant case such initially paired bosons can separate due to the multiphoton resonance. Further we calculate the quasienergy  spectrum and demonstrate that for the reduced interaction strength obeying $|u|>J$, the quasienergy is divided into two energy bands corresponding to the three paired states and the three unpaired states, respectively. Width of the energy gap between the two bands is proportional to the $|u|$ value.
The prediction on the CDT is confirmed by the Floquet quasienergy spectra in which the avoided level-crossings and new level-crossings near the collapse points are exhibited for the three unpaired states, due to the second-order corrections. While for the three paired states, a fine structure of quasienergy spectrum up to the second order is displayed by which we show the different level-crossings beyond the former collapse points.
The analytical results are very consistent with the direct numerical computations from the time-dependent Bose-Hubbard Hamiltonian for $|u|>10 J$. The second order results of long time scale could be conveniently applied to adiabatic manipulation \cite{Kral,PuH} of paired-particle tunneling in experiments \cite{Winkler,Folling}.

\section*{ACKNOWLEDGMENTS}
This work was supported by the NNSF of China under Grant No.
11175064 and No. 10905019, the Construct Program of the National Key Discipline,
the PCSIRTU of China (Grant No. IRT0964) and the Hunan Provincial NSF (Grant No. 11JJ7001).

\appendix*
\section{multiple-time-scale asymptotic analysis}
\renewcommand{\theequation}{A\arabic{equation}}

In the high-frequency regime, $\epsilon=J/\omega$ is a small positive parameter. Let $t'=\omega t$ be the
rescaling dimensionless time variable, we rewrite Eq. (4) as
\begin{eqnarray}
i\frac{da_1}{dt'}=&-&\sqrt{2}\epsilon a_4e^{[i\frac{U_0}{\omega}t'-i\varphi(t')]},\nonumber\\
i\frac{da_2}{dt'}=&-&\sqrt{2}\epsilon\Big[a_4 e^{[i\frac{U_0}{\omega}t'+i\varphi(t')]}+a_6 e^{[i\frac{U_0}{\omega}t'-i\varphi(t')]}\Big],\nonumber\\
i\frac{da_3}{dt'}=&-&\sqrt{2}\epsilon a_6e^{[i\frac{U_0}{\omega}t'+i\varphi(t')]},\nonumber\\
i\frac{da_4}{dt'}=&-&\epsilon\Big[\sqrt{2}a_1 e^{[-i\frac{U_0}{\omega}t'+i\varphi(t')]}+\sqrt{2}a_2 e^{[-i\frac{U_0}{\omega}t'-i\varphi(t')]}\nonumber\\&+&a_5 e^{[-i\varphi(t')]}\Big],\nonumber\\
i\frac{da_5}{dt'}=&-&\epsilon\Big[a_4 e^{i\varphi(t')}+a_6 e^{[-i\varphi(t')]}\Big],\nonumber\\
i\frac{da_6}{dt'}=&-&\epsilon\Big[\sqrt{2}a_2 e^{[-i\frac{U_0}{\omega}t'+i\varphi(t')]}+\sqrt{2}a_3 e^{[-i\frac{U_0}{\omega}t'-i\varphi(t')]}\nonumber\\&+&a_5 e^{i\varphi(t')}\Big],
\end{eqnarray}
with $\varphi(t')=\frac{\varepsilon}{\omega}\sin t'$. At first, we transform the independent variable $t'$  into the multiple-time-scale variables
$T_n=\epsilon^n t',n=0,1,2,...$, then replace the time derivatives
by the expansion
\begin{equation}
\frac{d}{dt'}=\partial_{T_0}+\epsilon\partial_{T_1}+\epsilon^2\partial_{T_2}+\cdot\cdot\cdot.
\end{equation}
At the same time, we expand $a_j(t') (j=1,2,...,6)$ as the power series
of $\epsilon$
\begin{equation}
a_j(t')=a_j^{(0)}(t')+\epsilon a_j^{(1)}(t')+\epsilon^2 a_j^{(2)}(t')+\cdot\cdot\cdot.
\end{equation}
Substituting Eqs. (A2) and (A3) into Eq. (A1), and collecting the terms of the same order,
we obtain a hierarchy of approximation equations of different orders in $\epsilon$. At
leading order $\epsilon^0$, one has
\begin{equation}
\frac{\partial a_j^{(0)}}{\partial T_0}=0,~~~a_j^{(0)}=A_j(T_1,T_2,\cdot\cdot\cdot),
\end{equation}
where the amplitudes $A_j(T_1,T_2,...)$ are functions of the slow
time variables $T_1,T_2,...$, but are independent of the fast
time variable $T_0$. At order $\epsilon$, one obtains the coupled equations
\begin{eqnarray}
i\frac{\partial a_1^{(1)}}{\partial T_0}=&-&i\frac{\partial A_1}{\partial T_1}-\sqrt{2}A_4e^{[i\frac{U_0}{\omega}T_0-i\varphi(T_0)]},\nonumber\\
i\frac{\partial a_2^{(1)}}{\partial T_0}=&-&i\frac{\partial A_2}{\partial T_1}-\sqrt{2}\Big[A_4e^{[i\frac{U_0}{\omega}T_0+i\varphi(T_0)]}\nonumber\\&+&A_6e^{[i\frac{U_0}{\omega}T_0-i\varphi(T_0)]}\Big],\nonumber\\
i\frac{\partial a_3^{(1)}}{\partial T_0}=&-&i\frac{\partial A_3}{\partial T_1}-\sqrt{2}A_6e^{[i\frac{U_0}{\omega}T_0+i\varphi(T_0)]},\nonumber\\
i\frac{\partial a_4^{(1)}}{\partial T_0}=&-&i\frac{\partial A_4}{\partial T_1}-\Big[\sqrt{2}A_1e^{[-i\frac{U_0}{\omega}T_0+i\varphi(T_0)]}\nonumber\\&+&\sqrt{2}A_2e^{[-i\frac{U_0}{\omega}T_0-i\varphi(T_0)]}+A_5
e^{[-i\varphi(T_0)]}\Big],\nonumber\\
i\frac{\partial a_5^{(1)}}{\partial T_0}=&-&i\frac{\partial A_5}{\partial T_1}-\Big[A_4e^{i\varphi(T_0)}+A_6e^{[-i\varphi(T_0)]}\Big],\nonumber\\
i\frac{\partial a_6^{(1)}}{\partial T_0}=&-&i\frac{\partial A_6}{\partial T_1}-\Big[\sqrt{2}A_2e^{[-i\frac{U_0}{\omega}T_0+i\varphi(T_0)]}\nonumber\\&+&\sqrt{2}A_3e^{[-i\frac{U_0}{\omega}T_0-i\varphi(T_0)]}+A_5
e^{i\varphi(T_0)}\Big].\nonumber\\
\end{eqnarray}
For the conveniences of our discussion, we simplify Eq. (A5) as $i\partial a_j^{(1)}/\partial T_0=-i\partial A_j/\partial T_1+G_j^{(1)}(T_0)$ for $j=1,2,...,6$.
To avoid the occurrence of secular growing terms in the solution $a_j^{(1)}$, the solvability condition \cite{Longhi2008, Longhi2012two}
\begin{equation}
i\frac{\partial A_j}{\partial T_1}=\overline{G_j^{(1)}(T_0)}
\end{equation}
must be satisfied, where the overline denotes the time average with respect to the fast time variable $T_0$, i.e., the dc component of the driving term $G_j^{(1)}(T_0)$. The amplitudes $a_j$ at order $\epsilon$ are given by
\begin{equation}
a_j^{(1)}=-i\int_0^{T_0}\Big[G_j^{(1)}(T_0)-\overline{G_j^{(1)}(T_0)}\ \Big]d\xi.
\end{equation}
Employing Eqs. (A5) and (A6), one gives
\begin{eqnarray}
&i\frac{\partial A_1}{\partial T_1}=i\frac{\partial A_2}{\partial T_1}=i\frac{\partial A_3}{\partial T_1}=0,\nonumber\\
&i\frac{\partial A_4}{\partial T_1}=i\frac{\partial A_6}{\partial T_1}=-\mathcal{J}_{0}(\frac{\varepsilon}{\omega})A_5,\nonumber\\
&i\frac{\partial A_5}{\partial T_1}=-\mathcal{J}_{0}(\frac{\varepsilon}{\omega})(A_4+A_6).
\end{eqnarray}
So the solutions of order $\epsilon$ read
\begin{eqnarray}
&&a_1^{(1)}=\sqrt{2}iA_4F_0(T_0),\nonumber\\
&&a_2^{(1)}=\sqrt{2}i[A_4F_1(T_0)+A_6F_0(T_0)],\nonumber\\
&&a_3^{(1)}=\sqrt{2}iA_6F_1(T_0),\nonumber\\
&&a_4^{(1)}=i[\sqrt{2}A_1F^*_0(T_0)+\sqrt{2}A_2F^*_1(T_0)+A_5F^*_2(T_0)],\nonumber\\
&&a_5^{(1)}=i[A_4F_2(T_0)+A_6F^*_2(T_0)],\nonumber\\
&&a_6^{(1)}=i[\sqrt{2}A_2F^*_0(T_0)+\sqrt{2}A_3F^*_1(T_0)+A_5F_2(T_0)],\nonumber\\
\end{eqnarray}
with
\begin{eqnarray}
&&F_0(T_0)=\sum\limits_{n'}\mathcal{J}_{n'}(\frac{\varepsilon}{\omega})\frac{\exp[i(\frac{U_0}{\omega}-n')T_0]-1}{i(\frac{U_0}{\omega}-n')},\nonumber\\
&&F_1(T_0)=\sum\limits_{n'}\mathcal{J}_{n'}(\frac{\varepsilon}{\omega})\frac{\exp[i(\frac{U_0}{\omega}+n')T_0]-1}{i(\frac{U_0}{\omega}+n')},\nonumber\\
&&F_2(T_0)=\sum\limits_{n'\neq 0}\mathcal{J}_{n'}(\frac{\varepsilon}{\omega})\frac{\exp(in'T_0)-1}{in'}.
\end{eqnarray}
We note that the probabilities to find the two strongly interacting bosons in the same wells are constants in time up to
the first-order time scale $T_1$ from Eq. (A8). Therefore, we need to consider the asymptotic analysis up to the order $\epsilon^2$.
Following the same procedure outlined above, one has $i\partial a_j^{(2)}/\partial T_0=-i\partial A_j/\partial T_2+G_j^{(2)}(T_0)$, with
\begin{eqnarray}
G_1^{(2)}=&&\sqrt{2}i\mathcal{J}_{0}(\frac{\varepsilon}{\omega})A_5F_0(T_0)-\sqrt{2}i(\sqrt{2}A_1F^*_0(T_0)\nonumber\\&&+\sqrt{2}A_2F^*_1(T_0)+A_5F^*_2(T_0))\exp(i\frac{U_0}{\omega}T_0\nonumber\\&&-i\varphi(T_0)),\nonumber\\
G_2^{(2)}=&&\sqrt{2}i\mathcal{J}_{0}(\frac{\varepsilon}{\omega})A_5(F_0(T_0)+F_1(T_0))-\nonumber\\&&\sqrt{2}i\Big[(\sqrt{2}A_1F^*_0(T_0)+A_5F_2^*(T_0)+\nonumber\\&&\sqrt{2}A_2F_1^*(T_0))
\exp(i\frac{U_0}{\omega}T_0+i\varphi(T_0))+\nonumber\\&&(\sqrt{2}A_2F_0^*(T_0)+\sqrt{2}A_3F_1^*(T_0)+\nonumber\\&&A_5F_2^*(T_0))\exp(i\frac{U_0}{\omega}T_0-i\varphi(T_0))\Big],\nonumber
\end{eqnarray}
\begin{eqnarray}
G_3^{(2)}=&&\sqrt{2}i\mathcal{J}_{0}(\frac{\varepsilon}{\omega})A_5F_1(T_0)-\sqrt{2}i(\sqrt{2}A_2F^*_0(T_0)\nonumber\\&&+\sqrt{2}A_3F^*_1(T_0)+A_5F_2(T_0))\exp(i\frac{U_0}{\omega}T_0\nonumber\\&&+i\varphi(T_0)),\nonumber\\
G_4^{(2)}=&&i\mathcal{J}_{0}(\frac{\varepsilon}{\omega})(A_4+A_6)F_2^*(T_0)-i\Big[2A_4F_0(T_0)\nonumber\\&&\exp(-i\frac{U_0}{\omega}T_0+i\varphi(T_0))+2(A_4F^*_1(T_0)\nonumber\\&&+A_6F_0^*(T_0))
\exp(-i\frac{U_0}{\omega}T_0-i\varphi(T_0))+\nonumber\\&&(A_4F_2(T_0)+A_6F_2^*(T_0))\exp(-i\varphi(T_0))\Big],\nonumber\\
G_5^{(2)}=&&i\mathcal{J}_{0}(\frac{\varepsilon}{\omega})A_5(F_2(T_0)+F_2^*(T_0))-\nonumber\\&&i\Big[(\sqrt{2}A_1F_0^*(T_0)+\sqrt{2}A_2F_1^*(T_0)+\nonumber\\&&A_5F_2^*(T_0))
\exp(i\varphi(T_0))+\nonumber\\&&(\sqrt{2}A_2F_0^*(T_0)+\sqrt{2}A_3F_1^*(T_0)+\nonumber \\&&
A_5F_2(T_0))\exp(-i\varphi(T_0))\Big],\nonumber\\
G_6^{(2)}=&&i\mathcal{J}_{0}(\frac{\varepsilon}{\omega})(A_4+A_6)F_2(T_0)-i\Big[\sqrt{2}(A_4F_1(T_0)\nonumber\\&&+A_6F_0(T_0))
\exp(-i\frac{U_0}{\omega}T_0+i\varphi(T_0))+\nonumber\\&&(A_4F_2(T_0)+A_6F_2^*(T_0))\exp(i\varphi(T_0))+\nonumber\\&&2A_6F_1(T_0)\exp(-i\frac{U_0}{\omega}T_0-i\varphi(T_0))\Big].
\end{eqnarray}
Then the solvability condition at order $\epsilon^2$ reads
\begin{eqnarray}
&&i\frac{\partial}{\partial T_2}A_1=\overline{G_1^{(2)}}=2(A_1\rho_1+A_2\rho_2),\nonumber\\
&&i\frac{\partial}{\partial T_2}A_2=\overline{G_2^{(2)}}=2(2A_1\rho_1+(A_1+A_3)\rho_2),\nonumber\\
&&i\frac{\partial}{\partial T_2}A_3=\overline{G_3^{(2)}}=2(A_3\rho_1+A_2\rho_2),\nonumber\\
&&i\frac{\partial}{\partial T_2}A_4=\overline{G_4^{(2)}}=-2(2A_4\rho_1+A_6\rho_2),\nonumber\\
&&i\frac{\partial}{\partial T_2}A_5=\overline{G_5^{(2)}}=0,\nonumber\\
&&i\frac{\partial}{\partial T_2}A_6=\overline{G_6^{(2)}}=-2(2A_6\rho_1+A_4\rho_2),
\end{eqnarray}
where we have set
\begin{equation}
\rho_1=\sum\limits_{n'}\frac{\mathcal{J}^2_{n'}(\frac{\varepsilon}{\omega})}{\frac{U_0}{\omega}+n'},~~~
\rho_2=\sum\limits_{n'}\frac{\mathcal{J}_{n'}(\frac{\varepsilon}{\omega})\mathcal{J}_{-n'}(\frac{\varepsilon}{\omega})}{\frac{U_0}{\omega}+n'}
\end{equation}
for $U_0/\omega+n'\ne 0$. Thus the evolution of the amplitudes $A_j$ up to the second-order long time is
given by
\begin{equation}
\frac{dA_j}{dt'}=(\frac{\partial}{\partial T_0}+\epsilon\frac{\partial}{\partial T_1}+\epsilon^2\frac{\partial}{\partial T_2})A_j,
\end{equation}
from Eq. (A3) and Eq. (A4). The corresponding probability amplitudes read
\begin{equation}
a_j(t')=A_j(t')+o(\epsilon)+o(\epsilon^2)+\cdot\cdot\cdot,
\end{equation}
in the high-frequency regime, where the high order small terms can be neglected. Substituting Eqs. (A4), (A8) and (A12) into Eq.(A14), we obtain the two sets of coupled equations,
Eqs. (6) and (7).

\end{document}